 \documentclass[final,3p,Ecuild]{elsarticle}

\usepackage{amssymb}
\usepackage{amsthm}
\usepackage{amsmath}
\usepackage{mathrsfs}
\usepackage{amsbsy}
\usepackage{color}

\usepackage{subfigure}

\usepackage{tikz}
\usetikzlibrary{shapes,arrows,positioning}

\usepackage{multirow}
\biboptions{sort&compress}

\usepackage{threeparttable}

\usepackage[colorlinks,linkcolor=red,anchorcolor=blue,citecolor=green]{hyperref} 

\DeclareMathOperator*{\argmax}{argmax}

\usepackage{easyReview}

 \usepackage{lineno}

\journal{arXiv}
\begin{document}
\begin{frontmatter}



\title{Response probability distribution estimation  of  expensive computer simulators: A Bayesian active learning perspective using Gaussian process regression}



\author[label1]{Chao Dang\corref{cor1}}
\ead{chao.dang@tu-dortmund.de}
\affiliation[label1]{organization={Chair for Reliability Engineering, TU Dortmund University},
	addressline={Leonhard-Euler-Str. 5}, 
	city={Dortmund 44227},
	country={Germany}}

	\author[label1]{Marcos A. Valdebenito}
	
	\author[label1]{Nataly  A.  Manque}
	
	\author[label2,label3]{Jun Xu}
	
	\author[label1]{Matthias G.R. Faes}

	\affiliation[label2]{organization={College of Civil Engineering, Hunan University},
		city={Changsha 410082},
		country={PR China}}

	\affiliation[label3]{organization={Key Lab on Damage Diagnosis for Engineering Structures of Hunan Province},
		city={Changsha 410082},
		country={PR China}}
	
	\cortext[cor1]{Corresponding author}

\begin{abstract}
Estimation of the response probability distributions of computer simulators in the presence of randomness is a crucial task in many fields. However, achieving this task with guaranteed accuracy remains an open computational challenge, especially for expensive-to-evaluate computer simulators. In this work, a Bayesian active learning perspective is presented to address the challenge, which is based on the use of the Gaussian process (GP) regression.  First,  estimation of the response probability distributions is conceptually interpreted as a Bayesian inference problem, as opposed to frequentist inference. This interpretation provides several important benefits: (1) it quantifies and propagates discretization error probabilistically; (2) it incorporates prior knowledge of the computer simulator, and (3) it enables the effective reduction of numerical uncertainty in the solution to a prescribed level. The conceptual Bayesian idea is then realized by using the GP regression, where we derive the posterior statistics of the response probability distributions in semi-analytical form and also provide a numerical solution scheme. Based on the practical Bayesian approach, a Bayesian active learning (BAL) method is further proposed for estimating the response probability distributions. In this context, the key contribution lies in the development of two crucial components for active learning, i.e., stopping criterion and learning function, by taking advantage of posterior statistics. It is empirically demonstrated by five numerical examples that the proposed BAL method can efficiently estimate the response probability distributions with  desired accuracy.

\end{abstract}




\end{frontmatter}


\section{Introduction}\label{}

Computer simulators are widely used across various fields of science and engineering to model, analyze, and predict the behavior of complex systems in the presence of randomness. For example, in physics, simulating quantum systems aids in understanding particle behavior, interactions, and probabilistic outcomes. In applied mechanics and engineering, finite element models are employed extensively for studying the performance of structures,  considering randomness in their internal structural properties and external operating conditions. In the latter context, typical research topics include: (1) Reliability analysis, which assesses the probability that a system produces an undesired response; (2) Statistical moment evaluation, focused on determining the statistical moments of the system's response; and (3) Probability distribution estimation, which involves estimating the probability distribution of the system response. Among these, a central problem is the estimation of the response probability distribution, such as the cumulative distribution function (CDF), complementary CDF (CCDF), and probability density function (PDF). This task is crucial because it provides a complete characterization of the uncertain system response, allowing a more comprehensive understanding of the underlying system behaviour under random influences without the need for costly and time-consuming physical experiments. However, it is often made difficult by the long running time of the computer simulator.

Over the past few decades, a variety of methods have been developed for estimating the response probability distributions of computer simulators. 
Commonly used methods  can be broadly classified into three types:  simulation  methods, statistical moment-based methods and surrogate-assisted methods. 
Often considered as  the most straightforward approach, simulation  methods estimate the response probability distribution by generating  numerous samples of the  response of interest. Examples of such methods include Monte Carlo simulation (MCS) \cite{kroese2013handbook}  and its various variants such as  stratified sampling \cite{helton2003latin,shields2015refined},    Latin hypercube sampling \cite{helton2003latin,shields2016generalization}  and quasi-MCS \cite{lemieux2009monte}.   In general, simulation methods are  less  or not  sensitive to the input dimensionality and non-linearity of the computer simulator under consideration. Nonetheless, they often suffer from slow convergence rates, necessitating a significant number of simulations.  This drawback becomes particularly pronounced when dealing with an expensive-to-evaluate computer simulator. As an alternative,  statistical moment-based methods approximate the probability distribution of the response of interest from knowledge of its estimated statistical moments through an assumed distribution model. Note that most, but not all, of the methods in this category were developed in the context of reliability analysis, but are equally applicable to the  topic of this work. A non-exhaustive list  includes high-order moments-based methods \cite{zhao2021structural,lee2009comparative,xu2019new,he2019maximum,zhou2020adaptive}, fractional moments-based maximum entropy methods \cite{zhang2013structural,xu2019adaptive} and fractional moments-based mixture distribution methods \cite{dang2020mixture,ding2023first}.  In addition,  recent novel  advances have been made from the perspective of  complex fractional moments \cite{xu2023generalized,song2024clustering,wan2024structural} and harmonic moments  \cite{xu2023harmonic,yu2023harmonic}. Nonetheless, a common criticism of statistical moment-based methods is that the numerical errors behind those probability distribution estimates are rarely known and remain challenging to derive.  Last but not least, surrogate-assisted methods have also been developed for response probability distribution estimation. The key idea is to  use  a  simplified model to substitute the original expensive computer simulator  based on a set of carefully selected input-output data points.  Representative methods include polynomial chaos expansions \cite{blatman2010adaptive,zeng2023projection}, Gaussian process (GP) regression  or Kriging \cite{oakley2002bayesian,wang2020novel,song2024bayesian,xiang2024multi}.  The interested reader is referred to \cite{su2024surrogate} for a comprehensive study of active learning based surrogates for estimating response probability distributions. 
It is shown that surrogate-assisted methods have the potential to reduce the computational burden if well designed.  In addition to the three types of methods mentioned above, it is worth noting that many other probability distribution estimation methods have been developed in specific field of stochastic dynamics, for example, path integrals \cite{kougioumtzogloupath},   globally-evolving-based generalized density evolution equation \cite{lyu2022unified}, probability density evolution method \cite{li2004probability,li2009stochastic}, direct probability integral method \cite{chen2019direct,tao2022fully}, to just name a few. It is noteworthy that the latter two methods can also be applied to stochastic static systems.  In summary, despite considerable efforts, it remains an open challenge to efficiently and accurately estimate the response probability distributions of expensive computer simulators.  In this context, most existing methods can become inefficient, inaccurate or even inapplicable depending on the particular characteristics of the problem at hand.

To address the research gap, this work aims to  present a Bayesian active learning perspective on  the response probability distribution estimation of expensive computer simulators  using GP regression.  The main contributions can be summarized as follows:
\begin{itemize}
	 \item The estimation of response probability distributions is conceptually interpreted  as a Bayesian inference problem.  This interpretation brings  several important benefits: (1) it provides a principled approach to quantifying and propagating discretisation error as a source of epistemic uncertainty in a probabilistic way through a computational pipeline; (2) it allows the incorporation of our prior knowledge  about the computer simulator into the estimation; and (3) it enables the effective reduction of  the numerical uncertainty in the solution of the response probability distribution to a prescribed level.

	 \item The conceptual Bayesian  idea is then realized by the virtue of the GP regression, an easy-to-use  Bayesian model to define a distribution over functions. A Gaussian process (GP) prior is first assigned to the computer simulator under consideration, and then conditioning the GP prior  on several computer simulator evaluations yields a posterior GP over the computer simulator. We  derive the posterior statistics of the response CDF, CCDF  and PDF in semi-analytical form and also provide the MCS solutions.  The developed Bayesian approach can be seen as an extension of the Bayesian approach for failure probability estimation reported in \cite{dang2021estimation,dang2022parallel,dang2022structural} and belongs to a probabilistic numerical method \cite{hennig2022probabilistic}.

	 \item The problem of response probability distribution estimation is finally framed in a Bayesian active learning setting. Specifically,   a Bayesian active learning method is proposed for estimating the response probability distributions, based on the above practical Bayesian approach. In this context, the key contribution is the development of two crucial components for active learning by making use of the posterior statistics: a stopping criterion and a learning function.
\end{itemize}

The rest of the paper is structured as follows.  Section \ref{sec:problem} presents the formulation of the problem to be solved in this work. The conceptual Bayesian framework for response probability distribution estimation  is given in Section \ref{sec:Bay_Inf}, followed by the practical one.  In Section \ref{sec:BAL}, we introduce the proposed Bayesian active learning method for  response probability distribution estimation. Five numerical examples are studied in Section \ref{sec:examples} to illustrate the proposed method.   Section \ref{sec:conclusions} concludes the paper with some final remarks.

\section{Problem formulation}\label{sec:problem}

Consider a physical model represented by a real-valued deterministic computer simulator $g: \mathbb{R}^d \mapsto \mathbb{R} $. The input to the model is  a vector of $d$ continuous random variables, i.e., $\boldsymbol{X}=[X_1, X_2, \cdots, X_d] \in \mathcal{X} \subseteq \mathbb{R}^d$, with the prescribed joint probability distribution function  $f_{\boldsymbol{X}}(\boldsymbol{x})$. As  a consequence, the response of the model is also a random variable denoted by $Y \in \mathcal{Y} \subseteq \mathbb{R}$, i.e., $Y = g(\boldsymbol{X})$.  The CDF of $Y$ is defined by:
\begin{equation}\label{eq:cdf}
	F_{Y}(y) = \mathbb{P}\left(Y \le y\right) = \mathbb{P}\left(g(\boldsymbol{X}) \le y\right) = \int_{\mathcal{X}} I_A(\boldsymbol{x})  f_{\boldsymbol{X}} (\boldsymbol{x}) \text{d} \boldsymbol{x},
\end{equation}
where $\mathbb{P}$ is the probability measure; $A=\left\{\boldsymbol{x} \in \mathcal{X} |g(\boldsymbol{x}) \le y \right\}$; $I_A(\boldsymbol{x})$ is the indicator function: $I_A(\boldsymbol{x}) = 1$ if $\boldsymbol{x} \in A$, and $I_A(\boldsymbol{x}) = 0$ otherwise. The CCDF of $Y$ is given by: 
\begin{equation}\label{eq:ccdf}
	\overline{F}_{Y}(y) = \mathbb{P}\left(Y > y\right) = \mathbb{P}\left(g(\boldsymbol{X}) > y\right) = \int_{\mathcal{X}} I_{\overline{A}}(\boldsymbol{x})  f_{\boldsymbol{X}} (\boldsymbol{x}) \text{d} \boldsymbol{x},
\end{equation}
where $\overline{A}$ is the complement  set of $A$, i.e.,  $\overline{A} = \left\{\boldsymbol{x} \in \mathcal{X} |g(\boldsymbol{x}) > y \right\}$. Note that $\overline{F}_{Y}(y) = 1 - F_{Y}(y)$ holds. In case we assume that $F_{Y}(y)$ is  almost everywhere differentiable, $Y$  admits a   PDF, which is expressed as:  
\begin{equation}\label{eq:pdf}
	f_{Y}(y) = \frac{d F_{Y} (y)}{d y} = \int_{\mathcal{X}} \delta (y-g(\boldsymbol{x})) f_{\boldsymbol{X}} (\boldsymbol{x}) \text{d} \boldsymbol{x} =  \int_{\mathcal{X}} \zeta_{B} (\boldsymbol{x}) f_{\boldsymbol{X}} (\boldsymbol{x}) \text{d} \boldsymbol{x} , 
\end{equation}
where $\delta(\cdot)$ is the Dirac delta function; $B = \left\{\boldsymbol{x} \in \mathcal{X}  |y=g(\boldsymbol{x})\right\}$;  $\zeta _B (\boldsymbol{x}) = \delta (y-g(\boldsymbol{x}))$.

The analytical solutions of Eqs. \eqref{eq:cdf} - \eqref{eq:pdf} are not available, except for some rather simple cases. Therefore, in practice, one has to resort to numerical methods. A numerical solution scheme often requires the computer simulator to be evaluated many times.  This can be computationally prohibitive if the computer simulator is expensive to evaluate, as is often the case.

\section{Bayesian inference about  response probability distributions}\label{sec:Bay_Inf}

This section treats the estimation of response probability distributions as a Bayesian inference problem, as opposed to frequentist inference. First, a conceptual Bayesian framework is created, which provides the philosophical foundations of this work. Second,  a practical Bayesian framework is developed based on the GP regression.

\subsection{Conceptual Bayesian framework}

\begin{figure}[htb]
	\centering
	\includegraphics[scale=0.45]{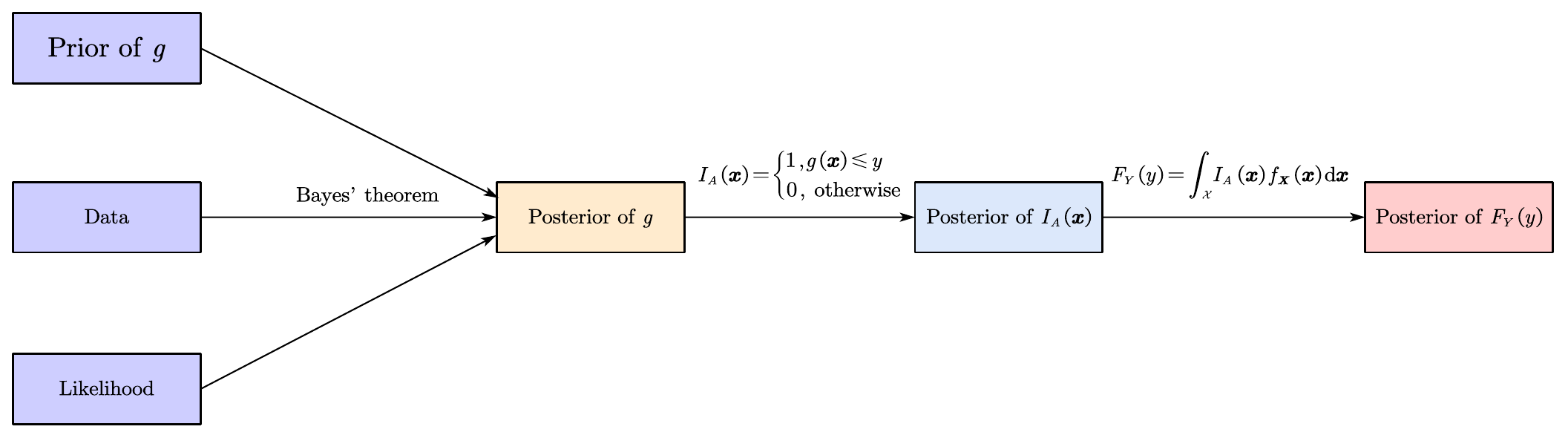}
	\caption{Conceptual illustration of the Bayesian inference about the response CDF.}\label{fig:concept_bay}
\end{figure}

As in many existing probabilistic numerical methods (e.g., Bayesian optimization \cite{jones1998efficient,garnett2023bayesian} and Bayesian quadrature \cite{o1991bayes,rasmussen2003bayesian,briol2019probabilistic}), the premise of a Bayesian inference treatment of the response probability distribution estimation is that the $g$ function should be thought of  as being random. This is because, even though  $g(\cdot)$ is by definition a deterministic mapping, it remains numerically unknown until we actually evaluate it. Furthermore, even when it can be evaluated, it is impractical to compute  $g(\cdot)$ at every possible location.  Once we admit that we do have epistemic uncertainty about $g(\cdot)$ due to discretization, it becomes natural to use a Bayesian approach to the problem of response probability distribution estimation. Take the estimation of the response CDF (Eq. \eqref{eq:cdf}) as an example. A Bayesian approach starts by placing a prior distribution on the $g$ function. Conditioning this prior on several evaluations of $g$ provides a posterior distribution over $g$ according to the Bayes' law. This posterior distribution subsequently induces a posterior distribution over the indicator function $I_A$, and ultimately, a posterior distribution over the desired response CDF $F_Y$.  The conceptual idea is illustrated in Fig. \ref{fig:concept_bay}. It should be noted that the response CCDF and PDF can be treated in a similar way.

The conceptual Bayesian framework offers several significant advantages, among which the following are particularly noteworthy:
\begin{itemize}
      \item  It provides a principled approach to quantifying and propagating the discretization error in a fully probabilistic manner through a computational pipeline. The Bayesian framework treats the discretization error as a source of epistemic uncertainty, enabling probabilistic analysis and reasoning. This allows for the systematic quantification and propagation of the discretization error through a computational pipeline. Note that the posterior distribution over the response  probability distribution reflects the fact that $g$ has been discretized. 
      
      \item   It allows the integration of our prior beliefs of the $g$ function into the estimation. Our prior knowledge about $g$, such as smoothness or periodicity and other properties, can be encoded by its prior distribution. This can often lead to faster convergence and reduced computational cost.

      \item  It facilitates the effective reduction of numerical uncertainty in the solution of the response probability distribution to a specified level. The well-quantified uncertainty can provide useful information for designing computer experiments in order to effectively reduce the numerical uncertainty to an accepted level.  
\end{itemize}

\subsection{Practical Bayesian framework}
\subsubsection{Prior distribution}

A stochastic process can be assigned to $g$ as a prior distribution to express our uncertainty associated with it. Among many possible options, this study adopts the widely-used GP prior such that:
\begin{equation}
	g_0 (\boldsymbol{x}) \sim \mathcal{GP}(m_{g_0} (\boldsymbol{x}), k_{g_0} (\boldsymbol{x}, \boldsymbol{x}^{\prime}) ),
\end{equation}
where $g_0$ denotes the prior distribution of $g$ before seeing any observations; $m_{g_0} (\boldsymbol{x}): \mathcal{X} \mapsto \mathbb{R}$ is prior mean function  and $k_{g_0} (\boldsymbol{x}, \boldsymbol{x}^{\prime}): \mathcal{X} \times \mathcal{X}  \mapsto \mathbb{R}$  is the prior covariance (also known as the kernel) function, i.e.,
\begin{equation}
	m_{g_0} (\boldsymbol{x}) = \mathbb{E}\left[ g_0 (\boldsymbol{x}) \right], 
\end{equation}
\begin{equation}
	k_{g_0} (\boldsymbol{x}, \boldsymbol{x}^{\prime}) = \mathbb{E}\left[ \left( g_0 (\boldsymbol{x}) - m_{g_0} (\boldsymbol{x})  \right)  \left( g_0 (\boldsymbol{x}^\prime) - m_{g_0} (\boldsymbol{x}^\prime)  \right) \right],
\end{equation}
where $\mathbb{E}$ is the expectation operator. The GP prior is fully specified by $m_{g_0} (\boldsymbol{x})$ and $k_{g_0} (\boldsymbol{x}, \boldsymbol{x}^{\prime})$. 
The prior mean function reflects the general trend of $g$, while the prior covariance function encodes key assumptions about the smoothness, periodicity, and other properties of $g$. Without loss of generality, we employ a constant prior mean and a Gaussian kernel:
\begin{equation}
	m_{g_0} (\boldsymbol{x}) = \beta,
\end{equation}
\begin{equation}
	k_{g_0} (\boldsymbol{x}, \boldsymbol{x}^{\prime}) = \sigma_0^2 \exp \left(- \frac{1}{2} (\boldsymbol{x}- \boldsymbol{x}^{ \prime})  \boldsymbol{\varSigma}^{-1} (\boldsymbol{x}^{} - \boldsymbol{x}^{ \prime}) ^{\top}   \right),
\end{equation}
where $\beta \in \mathbb{R}$; $\sigma_0>0$ is the  standard deviation of $g_0$; $\boldsymbol{\varSigma }=  \text{diag} \left\{l_1^2, l_2^2, \cdots, l_d^2\right\} $ is a diagonal matrix with $l_i>0$ being the characteristic length-scale   in the $i$-th dimension. The parameters collected in $\boldsymbol{\vartheta} = \left\{\beta ,\sigma_0, l_1,l_2, \cdots, l_d\right\}$ are known as hyper-parameters.

\subsubsection{Estimating hyper-parameters}

Assume that we now have a dataset $\boldsymbol{\mathcal{D}}=  \left\{ \boldsymbol{\mathcal{X}}, \boldsymbol{\mathcal{Y}} \right\} $, where $ \boldsymbol{\mathcal{X}} =  \left\{  \boldsymbol{x}^{(i)}\right\}_{i=1}^n$ is an $n$-by-$d$ matrix with its $i$-th row being $\boldsymbol{x}^{(i)}$, and $\boldsymbol{\mathcal{Y}} = \left\{y^{(i)}\right\}_{i=1}^{n_0}$ is an $n$-by-1 vector with $y^{(i)} = g (\boldsymbol{x}^{(i)})$. The hyper-parameters can be learned from data $\boldsymbol{\mathcal{D}}$ by maximizing the log marginal likelihood:
\begin{equation}
	\log p(\boldsymbol{\mathcal{Y}}| \boldsymbol{\mathcal{X}}, \boldsymbol{\vartheta} ) = 
	-\frac{1}{2}\left[ (\boldsymbol{\mathcal{Y}} -\beta)^{\top}\boldsymbol{K}_{g_0}^{-1}(\boldsymbol{\mathcal{Y}} -\beta )+\log |\boldsymbol{K}_{g_0}|+n\log (2\pi) \right],  
\end{equation} 
where $\boldsymbol{K}_{g_0}$ is an $n$-by-$n$ covariance matrix with its $(i,j)$-th entry is $\left[ \boldsymbol{K}_{g_0}  \right]_{i,j} =  	k_{g_0} (\boldsymbol{x}^{(i)}, \boldsymbol{x}^{(j)})$.

\subsubsection{Posterior statistics}

Conditioning the GP prior on the data $\boldsymbol{\mathcal{D}}$ gives the posterior distribution of $g$, which also follows a GP:
\begin{equation}
  g_n (\boldsymbol{x}) \sim   \mathcal{GP}(m_{g_n} (\boldsymbol{x}), k_{g_n} (\boldsymbol{x}, \boldsymbol{x}^{\prime}) ),
\end{equation}
where $g_n$ represents the posterior distribution of $g$ after seeing $n$ observations; $m_{g_n} (\boldsymbol{x}):  \mathcal{X} \mapsto \mathbb{R}$ is the posterior mean function, and $k_{g_n} (\boldsymbol{x}, \boldsymbol{x}^{\prime}): \mathcal{X} \times \mathcal{X}  \mapsto \mathbb{R}$  is the posterior covariance function: 
\begin{equation}
	m_{g_n}(\boldsymbol{x}) = m_{g_0}(\boldsymbol{x})+\boldsymbol{k}_{g_0}(\boldsymbol{x},\boldsymbol{\mathcal{X}} )^{\top}\boldsymbol{K}_{g_0}^{-1}\left( \boldsymbol{\mathcal{Y}} -\boldsymbol{m}_{g_0}(\boldsymbol{\mathcal{X}}) \right),
\end{equation}
\begin{equation}
	k_{g_n}(\boldsymbol{x},\boldsymbol{x}^{\prime})=k_{g_0}(\boldsymbol{x},\boldsymbol{x}^{\prime})-\boldsymbol{k}_{g_0}(\boldsymbol{x},\boldsymbol{\mathcal{X}} )^{\top}\boldsymbol{K}_{g_0}^{-1}\boldsymbol{k}_{g_0}(\boldsymbol{\mathcal{X}} ,\boldsymbol{x}^{\prime}),
\end{equation}
where $\boldsymbol{m}_{g_0}(\boldsymbol{\mathcal{X}})$ is an $n$-by-1 mean vector with its $i$-th element being $m(\boldsymbol{x}^{(i)})$; $\boldsymbol{k}_{g_0}(\boldsymbol{x},\boldsymbol{\mathcal{X}})$ is an $n$-by-1  covariance vector with its $i$-th element being $k_{g_0}(\boldsymbol{x},\boldsymbol{x}^{(i)})$; $\boldsymbol{k}_{g_0}(\boldsymbol{\mathcal{X}} ,\boldsymbol{x}^{\prime})$ is an $n$-by-1 vector with its $i$-th element being ${k}_{g_0}(\boldsymbol{x}^{(i)} ,\boldsymbol{x}^{\prime})$. For further detailed information on the above standard GP regression, please refer to  \cite{williams2006gaussian}.

The induced posterior distribution of $I_A$ conditional on $\boldsymbol{\mathcal{D}}$ follows a generalized  Bernoulli process (GBP):
\begin{equation}
	I_{A,n} (\boldsymbol{x}) \sim \mathcal{GBP} ( m_{I_{A,n}} (\boldsymbol{x}),   k_{I_{A,n}} (\boldsymbol{x}, \boldsymbol{x}^{\prime}) ),
\end{equation}
where $I_{A,n}$ denotes the posterior distribution of $I_A$ after seeing $n$ observations; $ m_{I_{A,n}} (\boldsymbol{x}): \mathcal{X} \mapsto \mathbb{R} $ and $k_{I_{A,n}} (\boldsymbol{x}, \boldsymbol{x}^{\prime}) : \mathcal{X} \times \mathcal{X}  \mapsto \mathbb{R}$  are the posterior mean and covariance functions respectively, which can be derived as follows:
\begin{equation}
	\begin{aligned}
   m_{I_{A,n}} (\boldsymbol{x}) = &  \mathbb{E}  \left[ I_{A,n} (\boldsymbol{x})     \right] \\  = & 1 \cdot  \mathbb{P} \left(  I_{A,n} (\boldsymbol{x}) = 1  \right)  + 0 \cdot  \mathbb{P} \left(  I_{A,n} (\boldsymbol{x}) = 0  \right) \\  = & \mathbb{P} \left( g_n (\boldsymbol{x}) \le y  \right) \\  = & \varPhi \left( \frac{y-m_{g_n}\left( \boldsymbol{x} \right)}{\sigma _{g_n}\left( \boldsymbol{x} \right)} \right),
   	\end{aligned}
\end{equation}
\begin{equation}
  \begin{aligned}
		k_{I_{A,n}} (\boldsymbol{x}, \boldsymbol{x}^{\prime}) =  &  	\mathbb{E}  \left[ \left(  I_{A,n} (\boldsymbol{x}) -m_{I_{A,n}} (\boldsymbol{x})  \right)  \left(  I_{A,n} (\boldsymbol{x}^\prime) -m_{I_{A,n}} (\boldsymbol{x}^\prime)  \right)   \right]  \\ = & \mathbb{E}  \left[   I_{A,n} (\boldsymbol{x})  I_{A,n} (\boldsymbol{x}^\prime) \right] - \mathbb{E} \left[   I_{A,n} (\boldsymbol{x})  \right] \mathbb{E}  \left[ I_{A,n} (\boldsymbol{x}^\prime) \right]  \\ = &  \mathbb{P} \left( g_n (\boldsymbol{x}) \le y, g_n (\boldsymbol{x}^\prime ) \le y   \right) - m_{I_{A,n}} (\boldsymbol{x}) m_{I_{A,n}} (\boldsymbol{x}^\prime) \\ = &  \varPhi _2\left( \left[ \begin{array}{c}
			y\\
			y\\
		\end{array} \right] ;\left[ \begin{array}{c}
			m_{g_n}\left( \boldsymbol{x} \right)\\
			m_{g_n}\left( \boldsymbol{x}^{\prime} \right)\\
		\end{array} \right] ,\left[ \begin{matrix}
			\sigma _{g_n}^{2}\left( \boldsymbol{x} \right)&		k_{g_n}(\boldsymbol{x},\boldsymbol{x}^{\prime})\\
			k_{g_n}(\boldsymbol{x}^{\prime},\boldsymbol{x})&		\sigma _{g_n}^{2}\left( \boldsymbol{x}^{\prime} \right)\\
		\end{matrix} \right] \right) -  \varPhi \left( \frac{y-m_{g_n}\left( \boldsymbol{x} \right)}{\sigma _{g_n}\left( \boldsymbol{x} \right)} \right) \varPhi \left( \frac{y-m_{g_n}\left( \boldsymbol{x}^{\prime} \right)}{\sigma _{g_n}\left( \boldsymbol{x}^{\prime} \right)} \right),
  \end{aligned}
\end{equation}
where  $\varPhi$ is the CDF of a standard normal variable; $\varPhi_2$ is the bivariate normal CDF; $\sigma _{g_n}\left( \boldsymbol{x} \right): \mathcal{X} \mapsto \mathbb{R}$ is the posterior standard deviation function of $g$, i.e., $\sigma _{g_n}\left( \boldsymbol{x} \right) = \sqrt{k_{g_n}(\boldsymbol{x},\boldsymbol{x})}$.  The posterior variance function of $I_A$, denoted by $\sigma^2_{I_{A,n}}\left( \boldsymbol{x} \right): \mathcal{X} \mapsto \mathbb{R}$, is given by:
\begin{equation}
	\sigma^2_{I_{A,n}}\left( \boldsymbol{x} \right) = \varPhi \left( \frac{y-m_{g_n}\left( \boldsymbol{x} \right)}{\sigma _{g_n}\left( \boldsymbol{x} \right)} \right) \varPhi \left(- \frac{y-m_{g_n}\left( \boldsymbol{x} \right)}{\sigma _{g_n}\left( \boldsymbol{x} \right)} \right).
\end{equation}

Similarly, the posterior distribution of $I_{\overline{A}}$ conditional on  $\boldsymbol{\mathcal{D}}$ also follows a GBP:
\begin{equation}
	I_{{\overline{A}},n} (\boldsymbol{x}) \sim \mathcal{GBP} ( m_{I_{{\overline{A}},n}} (\boldsymbol{x}),   k_{I_{{\overline{A}},n}} (\boldsymbol{x}, \boldsymbol{x}^{\prime})),
\end{equation}
where  $I_{{\overline{A}},n}$ denotes the posterior distribution of $I_{\overline{A}}$; $m_{I_{{\overline{A}},n}} (\boldsymbol{x}): \mathcal{X} \mapsto \mathbb{R}$ and $	k_{I_{{\overline{A}},n}} (\boldsymbol{x}, \boldsymbol{x}^{\prime}): \mathcal{X} \times \mathcal{X}  \mapsto \mathbb{R}$ are the posterior mean and covariance functions respectively, which are given by:
\begin{equation}
	m_{I_{{\overline{A}},n}} (\boldsymbol{x}) = \varPhi \left( - \frac{y-m_{g_n}\left( \boldsymbol{x} \right)}{\sigma _{g_n}\left( \boldsymbol{x} \right)} \right), 
\end{equation}
\begin{equation}
	k_{I_{{\overline{A}},n}} (\boldsymbol{x}, \boldsymbol{x}^{\prime}) =  	\varPhi _2\left( \left[ \begin{array}{c}
		y\\
		y\\
	\end{array} \right] ;\left[ \begin{array}{c}
		m_{g_n}\left( \boldsymbol{x} \right)\\
		m_{g_n}\left( \boldsymbol{x}^{\prime} \right)\\
	\end{array} \right] ,\left[ \begin{matrix}
		\sigma _{g_n}^{2}\left( \boldsymbol{x} \right)&		k_{g_n}(\boldsymbol{x},\boldsymbol{x}^{\prime})\\
		k_{g_n}(\boldsymbol{x}^{\prime},\boldsymbol{x})&		\sigma _{g_n}^{2}\left( \boldsymbol{x}^{\prime} \right)\\
	\end{matrix} \right] \right) -  \varPhi \left(\frac{y-m_{g_n}\left( \boldsymbol{x} \right)}{\sigma _{g_n}\left( \boldsymbol{x} \right)} \right) \varPhi \left( \frac{y-m_{g_n}\left( \boldsymbol{x}^{\prime} \right)}{\sigma _{g_n}\left( \boldsymbol{x}^{\prime} \right)} \right).
\end{equation}
The posterior variance function of $I_{\overline{A}}$, denoted by $\sigma^2_{I_{\overline{A},n}}\left( \boldsymbol{x} \right): \mathcal{X} \mapsto \mathbb{R}$, reads:
\begin{equation}
	\sigma^2_{I_{\overline{A},n}}\left( \boldsymbol{x} \right) = \varPhi \left( \frac{y-m_{g_n}\left( \boldsymbol{x} \right)}{\sigma _{g_n}\left( \boldsymbol{x} \right)} \right) \varPhi \left(- \frac{y-m_{g_n}\left( \boldsymbol{x} \right)}{\sigma _{g_n}\left( \boldsymbol{x} \right)} \right).
\end{equation}
We note that $k_{I_{A,n}} (\boldsymbol{x}, \boldsymbol{x}^{\prime})  = 	k_{I_{{\overline{A}},n}} (\boldsymbol{x}, \boldsymbol{x}^{\prime})$ and $\sigma^2_{I_{A,n}}\left( \boldsymbol{x} \right) = \sigma^2_{I_{\overline{A},n}}\left( \boldsymbol{x} \right)$ hold.

The posterior distribution of $\zeta_B$ conditional on $\boldsymbol{\mathcal{D}}$  follows a Dirac delta process (DDP): 
\begin{equation}
	\zeta_{B,n} (\boldsymbol{x})  \sim   \delta (y-g_n(\boldsymbol{x})),
\end{equation}
where $\zeta_{B,n}$ denotes the posterior distribution of $\zeta_B$.  The posterior mean and covariance functions of $\zeta_B$ can be derived as follows:
\begin{equation}
 \begin{aligned}
	m_{\zeta_{B,n}}(\boldsymbol{x}) = & \mathbb{E} \left[ \zeta_{B,n} (\boldsymbol{x})  \right] \\ = & \mathbb{E} \left[ \delta (y-g_n(\boldsymbol{x}))  \right] \\ = & \int_{-\infty}^{+\infty}  \delta (y-g_{n,i}(\boldsymbol{x})) f_{g_{n} \left( \boldsymbol{x} \right)} (g_{n,i}(\boldsymbol{x})) \text{d} g_{n,i} (\boldsymbol{x}) \\ = &  f_{g_{n} \left( \boldsymbol{x} \right)} (y) \\ = &   \frac{1}{\sigma _{g_n}\left( \boldsymbol{x} \right)}\phi \left( \frac{y-m_{g_n}\left( \boldsymbol{x} \right)}{\sigma _{g_n}\left( \boldsymbol{x} \right)} \right),
  \end{aligned}
\end{equation}
\begin{equation}
	 \begin{aligned}
	k_{\zeta_{B,n}}(\boldsymbol{x}, \boldsymbol{x}^\prime) =   &  	\mathbb{E}  \left[ \left(  \zeta_{B,n} (\boldsymbol{x}) -m_{\zeta_{B,n}} (\boldsymbol{x})  \right)  \left(  \zeta_{B,n} (\boldsymbol{x}^\prime) -m_{\zeta_{B,n}} (\boldsymbol{x}^\prime)  \right)   \right]  \\ = & \mathbb{E}  \left[   \zeta_{B,n} (\boldsymbol{x})  \zeta_{B,n} (\boldsymbol{x}^\prime) \right] - \mathbb{E}  \left[   \zeta_{B,n} (\boldsymbol{x})  \right] \mathbb{E} \left[ \zeta_{B,n}  (\boldsymbol{x}^\prime) \right] \\ = & \mathbb{E} \left[ \delta (y-g_n(\boldsymbol{x})) \delta (y-g_n(\boldsymbol{x}^\prime))  \right] - m_{\zeta_{B,n}}(\boldsymbol{x}) m_{\zeta_{B,n}}(\boldsymbol{x}^\prime) \\ = &  \int_{-\infty}^{+\infty} \int_{-\infty}^{+\infty}  \delta (y-g_{n,i}(\boldsymbol{x})) \delta (y-g_{n,j}(\boldsymbol{x}^\prime)) f_{g_{n}(\boldsymbol{x}),g_{n}(\boldsymbol{x}^\prime)} \left(  g_{n,i}(\boldsymbol{x}) , g_{n,j}(\boldsymbol{x}^\prime) \right) \text{d} g_{n,i} (\boldsymbol{x}) \text{d} g_{n,i} (\boldsymbol{x}^\prime) \\& - m_{\zeta_{B,n}}(\boldsymbol{x}) m_{\zeta_{B,n}}(\boldsymbol{x}^\prime) \\ = & f_{g_{n}(\boldsymbol{x}),g_{n}(\boldsymbol{x}^\prime)} \left(  y , y \right)   - m_{\zeta_{B,n}}(\boldsymbol{x}) m_{\zeta_{B,n}}(\boldsymbol{x}^\prime) \\ = &  \phi _2\left( \left[ \begin{array}{c}
		y\\
		y\\
	\end{array} \right] ;\left[ \begin{array}{c}
		m_{g_n}\left( \boldsymbol{x} \right)\\
		m_{g_n}\left( \boldsymbol{x}^{\prime} \right)\\
	\end{array} \right] ,\left[ \begin{matrix}
		\sigma _{g_n}^{2}\left( \boldsymbol{x} \right)&		k_{g_n}(\boldsymbol{x},\boldsymbol{x}^{\prime})\\
		k_{g_n}(\boldsymbol{x}^{\prime},\boldsymbol{x})&		\sigma _{g_n}^{2}\left( \boldsymbol{x}^{\prime} \right)\\
	\end{matrix} \right] \right) \\ & -  \frac{1}{\sigma _{g_n}\left( \boldsymbol{x} \right)}\phi \left( \frac{y-m_{g_n}\left( \boldsymbol{x} \right)}{\sigma _{g_n}\left( \boldsymbol{x} \right)} \right) \frac{1}{\sigma _{g_n}\left( \boldsymbol{x}^\prime \right)}\phi \left( \frac{y-m_{g_n}\left( \boldsymbol{x}^\prime \right)}{\sigma _{g_n}\left( \boldsymbol{x}^\prime \right)} \right), 
	  \end{aligned}
\end{equation}
where $g_{n,i}$ and  $g_{n,j}$  represents two realizations of  $g_n$; $f_{g_{n} \left( \boldsymbol{x} \right)}$ is the PDF of $g_{n} \left( \boldsymbol{x} \right)$; $f_{g_{n}(\boldsymbol{x}),g_{n}(\boldsymbol{x}^\prime)}$ is the joint PDF of $g_{n}(\boldsymbol{x})$ and $g_{n}(\boldsymbol{x}^\prime)$; $\phi_2$ is the bi-variate normal PDF. The posterior variance function of $\zeta_B$ is expressed as:
\begin{equation}
	\sigma^2_{\zeta_{B,n}}(\boldsymbol{x}) = \left[ \delta (0) - \frac{1}{\sigma _{g_n}\left( \boldsymbol{x} \right)}\phi \left( \frac{y-m_{g_n}\left( \boldsymbol{x} \right)}{\sigma _{g_n}\left( \boldsymbol{x} \right)} \right) \right] \frac{1}{\sigma _{g_n}\left( \boldsymbol{x} \right)}\phi \left( \frac{y-m_{g_n}\left( \boldsymbol{x} \right)}{\sigma _{g_n}\left( \boldsymbol{x} \right)} \right). 
\end{equation}
As $\delta(0)=\infty$, the posterior variance  function of $\zeta_B$ does not exist.

The posterior distribution (denoted by $F_{Y,n}$) of $F_{Y}$ can be generated by considering the push-forward of $I_{A,n}$ by the integration operator. However,  the exact type of this distribution  is not known. Fortunately, we can derive the posterior mean and variance functions by applying  Fubini's theorem as follows:
\begin{equation}\label{eq:post_mean_CDF}
	\begin{aligned}
			 m_{F_{Y,n}}\left( y \right) = & \mathbb{E} \left[ F_{Y,n} (y)  \right] \\ = &  \mathbb{E} \left[  \int_{\mathcal{X}} I_{A,n}(\boldsymbol{x})  f_{\boldsymbol{X}} (\boldsymbol{x}) \text{d} \boldsymbol{x}   \right] \\ = & \int_{\mathcal{X}} \mathbb{E} \left[ I_{A,n}(\boldsymbol{x}) \right] f_{\boldsymbol{X}} (\boldsymbol{x}) \text{d} \boldsymbol{x}  \\ = &  \int_{\mathcal{X}}{\varPhi \left( \frac{y-m_{g_n}\left( \boldsymbol{x} \right)}{\sigma _{g_n}\left( \boldsymbol{x} \right)} \right) f_{\boldsymbol{X}}\left( \boldsymbol{x} \right) \mathrm{d}\boldsymbol{x}},
	\end{aligned}
\end{equation}
\begin{equation}
	\begin{aligned}
		\sigma^2_{F_{Y,n}}\left( y \right) = & \mathbb{V} \left[  {F_{Y,n}}\left( y \right)  \right] \\ = &  \mathbb{E} \left[ \left( {F_{Y,n}}\left( y \right) -  m_{F_{Y,n}}\left( y \right) \right)^2 \right]  \\ = &  \mathbb{E} \left[ \left( \int_{\mathcal{X}}{I_{A,n}}(\boldsymbol{x})f_{\boldsymbol{X}}(\boldsymbol{x})\mathrm{d}\boldsymbol{x}-\int_{\mathcal{X}}{\mathbb{E}}\left[ I_{A,n}(\boldsymbol{x}) \right] f_{\boldsymbol{X}}(\boldsymbol{x})\mathrm{d}\boldsymbol{x} \right) ^2 \right]   \\=  &   \mathbb{E} \left[ \left( \int_{\mathcal{X}}{\left( I_{A,n}(\boldsymbol{x})-\mathbb{E} \left[ I_{A,n}(\boldsymbol{x}) \right] \right) f_{\boldsymbol{X}}(\boldsymbol{x})\mathrm{d}\boldsymbol{x}} \right) ^2 \right]  \\ = & \mathbb{E} \left[ \left( \int_{\mathcal{X}}{\left[ I_{A,n}(\boldsymbol{x})-\mathbb{E} \left[ I_{A,n}(\boldsymbol{x}) \right] \right]}f_{\boldsymbol{X}}(\boldsymbol{x})\mathrm{d}\boldsymbol{x} \right) \left( \int_{\mathcal{X}}{\left[ I_{A,n}(\boldsymbol{x}^{\prime})-\mathbb{E} \left[ I_{A,n}(\boldsymbol{x}^{\prime}) \right] \right]}f_{\boldsymbol{X}}(\boldsymbol{x}^{\prime})\mathrm{d}\boldsymbol{x}^{\prime} \right) \right]  \\ = & \mathbb{E} \left[ \left( \int_{\mathcal{X}}{\int_{\mathcal{X}}{\left( I_{A,n}(\boldsymbol{x})-\mathbb{E} \left[ I_{A,n}(\boldsymbol{x}) \right] \right) \left( I_{A,n}(\boldsymbol{x}^{\prime})-\mathbb{E} \left[ I_{A,n}(\boldsymbol{x}^{\prime}) \right] \right)}f_{\boldsymbol{X}}(\boldsymbol{x})f_{\boldsymbol{X}}(\boldsymbol{x}^{\prime})\mathrm{d}\boldsymbol{x}\mathrm{d}\boldsymbol{x}^{\prime}} \right) \right] \\ = & \int_{\mathcal{X}}{\int_{\mathcal{X}}{\mathbb{E} \left[ \left( I_{A,n}(\boldsymbol{x})-\mathbb{E} \left[ I_{A,n}(\boldsymbol{x}) \right] \right) \left( I_{A,n}(\boldsymbol{x}^{\prime})-\mathbb{E} \left[ I_{A,n}(\boldsymbol{x}^{\prime}) \right] \right) \right]}f_{\boldsymbol{X}}(\boldsymbol{x})f_{\boldsymbol{X}}(\boldsymbol{x}^{\prime})\mathrm{d}\boldsymbol{x}\mathrm{d}\boldsymbol{x}^{\prime}} \\= & \int_{\mathcal{X}}{\int_{\mathcal{X}}{k_{I_{A,n}}(\boldsymbol{x},\boldsymbol{x}^{\prime})}f_{\boldsymbol{X}}(\boldsymbol{x})f_{\boldsymbol{X}}(\boldsymbol{x}^{\prime})\mathrm{d}\boldsymbol{x}\mathrm{d}\boldsymbol{x}^{\prime}} \\ = & \int_{\mathcal{X}}{\int_{\mathcal{X}}{\left\{ \begin{array}{c}
					\varPhi _2\left( \left[ \begin{array}{c}
						y\\
						y\\
					\end{array} \right] ;\left[ \begin{array}{c}
						m_{g_n}\left( \boldsymbol{x} \right)\\
						m_{g_n}\left( \boldsymbol{x}^{\prime} \right)\\
					\end{array} \right] ,\left[ \begin{matrix}
						\sigma _{g_n}^{2}\left( \boldsymbol{x} \right)&		k_{g_n}(\boldsymbol{x},\boldsymbol{x}^{\prime})\\
						k_{g_n}(\boldsymbol{x}^{\prime},\boldsymbol{x})&		\sigma _{g_n}^{2}\left( \boldsymbol{x}^{\prime} \right)\\
					\end{matrix} \right] \right)\\
					-\varPhi \left( \frac{y-m_{g_n}\left( \boldsymbol{x} \right)}{\sigma _{g_n}\left( \boldsymbol{x} \right)} \right) \varPhi \left( \frac{y-m_{g_n}\left( \boldsymbol{x}^{\prime} \right)}{\sigma _{g_n}\left( \boldsymbol{x}^{\prime} \right)} \right)\\
				\end{array} \right\} f_{\boldsymbol{X}}(\boldsymbol{x})f_{\boldsymbol{X}}(\boldsymbol{x}^{\prime} )\mathrm{d}\boldsymbol{x}\mathrm{d}\boldsymbol{x}^{\prime}}},
	\end{aligned}
\end{equation}
where $\mathbb{V}$ is the variance operator. To avoid the complexity of calculating the posterior variance, an upper bound on $ \sigma^2_{F_{Y,n}}\left( y \right) $ is derived using the Cauchy-Schwarz inequality (i.e., $k_{I_{A,n}}(\boldsymbol{x},\boldsymbol{x}^{\prime}) \le \sigma_{I_{{A},n}}\left( \boldsymbol{x} \right)  \sigma_{I_{{A},n}}\left( \boldsymbol{x}^\prime \right) $):
\begin{equation}\label{eq:up_post_variance_CDF}
	\begin{aligned}
	\overline{\sigma}^2_{F_{Y,n}}\left( y \right) =  & \int_{\mathcal{X}}{\int_{\mathcal{X}}{\sigma _{I_{{A},n}}\left( \boldsymbol{x} \right) \sigma _{I_{{A},n}}\left( \boldsymbol{x}^{\prime} \right) f_{\boldsymbol{X}}(\boldsymbol{x})f_{\boldsymbol{X}}(\boldsymbol{x}^{\prime})\mathrm{d}\boldsymbol{x}\mathrm{d}\boldsymbol{x}^{\prime}}} \\ = & \left( \int_{\mathcal{X}}{\sigma _{I_{{A},n}}\left( \boldsymbol{x} \right) f_{\boldsymbol{X}}(\boldsymbol{x})\mathrm{d}\boldsymbol{x}} \right) ^2 \\ = &  \left( \int_{\mathcal{X}}{\sqrt{\varPhi \left( \frac{y-m_{g_n}\left( \boldsymbol{x} \right)}{\sigma _{g_n}\left( \boldsymbol{x} \right)} \right) \varPhi \left( -\frac{y-m_{g_n}\left( \boldsymbol{x} \right)}{\sigma _{g_n}\left( \boldsymbol{x} \right)} \right)}f_{\boldsymbol{X}}(\boldsymbol{x})\mathrm{d}\boldsymbol{x}} \right) ^2.
	\end{aligned}
\end{equation}

Analogously, the posterior mean and variance functions of  $\overline{F}_{Y}$ are given by:
\begin{equation}\label{eq:post_mean_CCDF}
	m_{\overline{F}_{Y,n}}\left( y \right) = \int_{\mathcal{X}}{\varPhi \left(- \frac{y-m_{g_n}\left( \boldsymbol{x} \right)}{\sigma _{g_n}\left( \boldsymbol{x} \right)} \right) f_{\boldsymbol{X}}\left( \boldsymbol{x} \right) \mathrm{d}\boldsymbol{x}},
\end{equation}
\begin{equation}
	\sigma^2_{\overline{F}_{Y,n}}\left( y \right) = \int_{\mathcal{X}}{\int_{\mathcal{X}}{\left\{ \begin{array}{c}
				\varPhi _2\left( \left[ \begin{array}{c}
					y\\
					y\\
				\end{array} \right] ;\left[ \begin{array}{c}
					m_{g_n}\left( \boldsymbol{x} \right)\\
					m_{g_n}\left( \boldsymbol{x}^{\prime} \right)\\
				\end{array} \right] ,\left[ \begin{matrix}
					\sigma _{g_n}^{2}\left( \boldsymbol{x} \right)&		k_{g_n}(\boldsymbol{x},\boldsymbol{x}^{\prime})\\
					k_{g_n}(\boldsymbol{x}^{\prime},\boldsymbol{x})&		\sigma _{g_n}^{2}\left( \boldsymbol{x}^{\prime} \right)\\
				\end{matrix} \right] \right)\\
				-\varPhi \left( \frac{y-m_{g_n}\left( \boldsymbol{x} \right)}{\sigma _{g_n}\left( \boldsymbol{x} \right)} \right) \varPhi \left( \frac{y-m_{g_n}\left( \boldsymbol{x}^{\prime} \right)}{\sigma _{g_n}\left( \boldsymbol{x}^{\prime} \right)} \right)\\
			\end{array} \right\} f_{\boldsymbol{X}}(\boldsymbol{x})f_{\boldsymbol{X}}(\boldsymbol{x}^{\prime} )\mathrm{d}\boldsymbol{x}\mathrm{d}\boldsymbol{x}^{\prime}}}.
\end{equation}
Furthermore, an upper bound on $\sigma^2_{\overline{F}_{Y,n}}\left( y \right)$ is available:
\begin{equation}\label{eq:up_post_variance_CCDF}
	\overline{\sigma}^2_{\overline{F}_{Y,n}}\left( y \right) =  \left( \int_{\mathcal{X}}{\sqrt{\varPhi \left( \frac{y-m_{g_n}\left( \boldsymbol{x} \right)}{\sigma _{g_n}\left( \boldsymbol{x} \right)} \right) \varPhi \left( -\frac{y-m_{g_n}\left( \boldsymbol{x} \right)}{\sigma _{g_n}\left( \boldsymbol{x} \right)} \right)}f_{\boldsymbol{X}}(\boldsymbol{x})\mathrm{d}\boldsymbol{x}} \right) ^2.
\end{equation}
It should be noted that there exist $\sigma^2_{\overline{F}_{Y,n}}\left( y \right) = \sigma^2_{{F}_{Y,n}}\left( y \right)$ and $\overline{\sigma}^2_{\overline{F}_{Y,n}}\left( y \right) = \overline{\sigma}^2_{{F}_{Y,n}}\left( y \right)$.

Likewise, we can obtain the posterior mean and variance functions of $f_{Y}$: 
\begin{equation}
	m_{f_{Y,n}}\left( y \right) =\int_{\mathcal{X}}{\frac{1}{\sigma _{g_n}\left( \boldsymbol{x} \right)}\phi \left( \frac{y-m_{g_n}\left( \boldsymbol{x} \right)}{\sigma _{g_n}\left( \boldsymbol{x} \right)} \right) f_{\boldsymbol{X}}\left( \boldsymbol{x} \right) \mathrm{d}\boldsymbol{x}},
\end{equation}
\begin{equation}
	\sigma^2_{f_{Y,n}}\left( y \right) = \int_{\mathcal{X}}{\int_{\mathcal{X}}{\left\{ \begin{array}{c}
				\phi_2\left( \left[ \begin{array}{c}
					y\\
					y\\
				\end{array} \right] ;\left[ \begin{array}{c}
					m_{g_n}\left( \boldsymbol{x} \right)\\
					m_{g_n}\left( \boldsymbol{x}^{\prime} \right)\\
				\end{array} \right] ,\left[ \begin{matrix}
					\sigma _{g_n}^{2}\left( \boldsymbol{x} \right)&		k_{g_n}(\boldsymbol{x},\boldsymbol{x}^{\prime})\\
					k_{g_n}(\boldsymbol{x}^{\prime},\boldsymbol{x})&		\sigma _{g_n}^{2}\left( \boldsymbol{x}^{\prime} \right)\\
				\end{matrix} \right] \right)\\
				-\frac{1}{\sigma _{g_n}\left( \boldsymbol{x} \right)}\phi \left( \frac{y-m_{g_n}\left( \boldsymbol{x} \right)}{\sigma _{g_n}\left( \boldsymbol{x} \right)} \right) \frac{1}{\sigma _{g_n}\left( \boldsymbol{x}^\prime \right)}\phi \left( \frac{y-m_{g_n}\left( \boldsymbol{x}^\prime \right)}{\sigma _{g_n}\left( \boldsymbol{x}^\prime \right)} \right)\\
			\end{array} \right\} f_{\boldsymbol{X}}(\boldsymbol{x})f_{\boldsymbol{X}}(\boldsymbol{x}^{\prime} )\mathrm{d}\boldsymbol{x}\mathrm{d}\boldsymbol{x}^{\prime}}}.
\end{equation}
In addition, an upper bound on  $\sigma^2_{f_{Y,n}}\left( y \right)$ can also be derived:
\begin{equation}\label{eq:upper_var_pdf}
	\overline{\sigma}^2_{f_{Y,n}}\left( y \right) =  \left( \int_{\mathcal{X}}{\sqrt{\left[ \delta (0) - \frac{1}{\sigma _{g_n}\left( \boldsymbol{x} \right)}\phi \left( \frac{y-m_{g_n}\left( \boldsymbol{x} \right)}{\sigma _{g_n}\left( \boldsymbol{x} \right)} \right) \right] \frac{1}{\sigma _{g_n}\left( \boldsymbol{x} \right)}\phi \left( \frac{y-m_{g_n}\left( \boldsymbol{x} \right)}{\sigma _{g_n}\left( \boldsymbol{x} \right)} \right)}f_{\boldsymbol{X}}(\boldsymbol{x})\mathrm{d}\boldsymbol{x}} \right) ^2. 
\end{equation}
Eq. \eqref{eq:upper_var_pdf} implies that the upper limit of $\sigma^2_{f_{Y,n}}\left( y \right)$ goes to infinity, so it is not meaningful.

The posterior mean functions of the response CDF, CCDF and PDF  can naturally provide point estimates, while the posterior variance functions (or their upper bound functions, if they exist) can serve as measures of uncertainty for the response probability distributions.

\subsubsection{Numerical treatment of posterior statistics}

Note that all posterior statistics of $F_{Y}$, $\overline{F}_{Y}$ and $f_{Y}$ are not analytically tractable (if exist) and thus require numerical treatment. In this study, Monte Carlo simulation (MCS) is employed. Taking the posterior statistics of $F_{Y}$ as an example, the estimators of $m_{F_{Y,n}}\left( y \right)$ and $\overline{\sigma}_{F_{Y,n}}\left( y \right)$ are given by:
\begin{equation}
	\hat{m}_{F_{Y,n}}\left( y \right) = \frac{1}{N} \sum_{j=1}^{N} \varPhi \left( \frac{y-m_{g_n}\left( \boldsymbol{x}^{(j)} \right)}{\sigma _{g_n}\left( \boldsymbol{x}^{(j)} \right)} \right),
\end{equation}
\begin{equation}
	\hat{\overline{\sigma}}_{F_{Y,n}}\left( y \right) = \frac{1}{N}\sum_{j=1}^N{\sqrt{\varPhi \left( \frac{y-m_{g_n}\left( \boldsymbol{x}^{(j)} \right)}{\sigma _{g_n}\left( \boldsymbol{x}^{(j)} \right)} \right) \varPhi \left( -\frac{y-m_{g_n}\left( \boldsymbol{x}^{(j)} \right)}{\sigma _{g_n}\left( \boldsymbol{x}^{(j)} \right)} \right)}},
\end{equation}
 where $\left\{\boldsymbol{x}^{(j)}\right\}_{j=1}^N$  is a set of $N$ random samples generated according to $f_{\boldsymbol{X}}(\boldsymbol{x})$. The  variances of $\hat{m}_{F_{Y,n}}\left( y \right)$ and $\hat{\overline{\sigma}}_{F_{Y,n}}\left( y \right)$ can be expressed as:
 \begin{equation}
 	\mathbb{V} \left[  \hat{m}_{F_{Y,n}}\left( y \right)  \right] = \frac{1}{N(N-1)} \sum_{j=1}^{N} \left[   \varPhi \left( \frac{y-m_{g_n}\left( \boldsymbol{x}^{(j)} \right)}{\sigma _{g_n}\left( \boldsymbol{x}^{(j)} \right)} \right) - \hat{m}_{F_{Y,n}}\left( y \right) \right]^2,
 \end{equation}
 \begin{equation}
  	\mathbb{V} \left[ \hat{\overline{\sigma}}_{F_{Y,n}}\left( y \right)  \right] =   \frac{1}{N(N-1)} \sum_{j=1}^{N} \left[   \sqrt{\varPhi \left( \frac{y-m_{g_n}\left( \boldsymbol{x}^{(j)} \right)}{\sigma _{g_n}\left( \boldsymbol{x}^{(j)} \right)} \right) \varPhi \left( -\frac{y-m_{g_n}\left( \boldsymbol{x}^{(j)} \right)}{\sigma _{g_n}\left( \boldsymbol{x}^{(j)} \right)} \right)} - \hat{\overline{\sigma}}_{F_{Y,n}}\left( y \right) \right]^2 .
 \end{equation}
  
Following the practical Bayesian framework described above, one can obtain point estimates for the response probability distributions, along with their associated uncertainty measure estimates, given the data  $\boldsymbol{\mathcal{D}}$.  The design of computer experiments thus is important for the accuracy and efficiency, which  involves the determination of the number and locations of the input samples.

\section{Bayesian active learning of response probability distributions}\label{sec:BAL}

In this section, a Bayesian active learning method is developed based on the practical Bayesian framework for estimating response probability distributions. This implies that the response probability distributions are estimated in an iterative way until a certain criterion is fulfilled. Since an upper bound on the posterior variance function for the response PDF does not exist, we will focus on the response CDF and CCDF in the active learning process, with the response PDF being obtained as a by-product at the end.

\subsection{Stopping criterion}

One of the 	critical challenges in Bayesian active learning is determining  when to stop  the iterative process, known as stopping criterion.   In general, formulating a stopping criterion might  depend on several considerations, such as the primary goal, available resources, and other factors. Hereto, the accuracy of the response CDF and CCDF is of great interest.

A natural measure of the accuracy of the response CDF is its posterior coefficient of variation (CoV) function.  However,  such a measure involves the posterior variance function of the response CDF,  which can be computationally demanding.  Alternatively, we use the upper bound of the posterior CoV function of the response CDF as a measure of its accuracy, which is defined as:
\begin{equation}
  \overline{\mathrm{CoV}} _{F_{Y,n}} (y) =  \frac{\overline{\sigma}_{F_{Y,n}}\left( y \right) }{m_{F_{Y,n}}\left( y \right)} ,
\end{equation}
where $m_{F_{Y,n}}\left( y \right)$ is given in Eq. \eqref{eq:post_mean_CDF}  and $ \overline{\sigma}_{F_{Y,n}}\left( y \right) $ is given in Eq. \eqref{eq:up_post_variance_CDF}.  Likewise, the upper bound of the posterior CoV function of the response CCDF is used as a measure of its accuracy:
\begin{equation}
	\overline{\mathrm{CoV}} _{\overline{F}_{Y,n}}  (y) = \frac{\overline{\sigma}_{\overline{F}_{Y,n}}\left( y \right)}{m_{\overline{F}_{Y,n}}\left( y \right)},
\end{equation}
where $m_{\overline{F}_{Y,n}}\left( y \right)$ is given in Eq. \eqref{eq:post_mean_CCDF} and $\overline{\sigma}_{\overline{F}_{Y,n}}\left( y \right)$ is given in Eq. \eqref{eq:up_post_variance_CCDF}.

In order to ensure the accuracy of both the response CDF and CCDF, we introduce the following stopping criterion in this study:
\begin{equation}\label{eq:stopping_criterion}
	   \max_{y \in \mathcal{Y}} H_n(y)  < \epsilon,
\end{equation}
where $H_n(y)= \max   \left( \overline{\mathrm{CoV}} _{F_{Y,n}} (y), \overline{\mathrm{CoV}} _{\overline{F}_{Y,n}}  (y)  \right) $; $\epsilon$ is a user-specified threshold.  This stopping criterion means that the iterative process stops as soon as the upper bounds of the posterior CoV  functions of both the response CDF and CCDF for any $y \in \mathcal{Y}$ are less than a predefined threshold $\epsilon$.

\subsection{Learning function}

Another critical component in Bayesian active learning is the mechanism or strategy used to select the most informative data points to evaluate next, known as the learning (or acquisition) function. This function comes into play  when the stopping criterion (Ineq. \eqref{eq:stopping_criterion}) is not met.  The sought learning function should be able to suggest promising points which, once evaluated, may reduce the value of the left-hand side term of Ineq. \eqref{eq:stopping_criterion}.

First, we identify a critical location at which $H_n(y)$ has the largest value:
\begin{equation}
	  y^{\star} = \argmax_{y \in \mathcal{Y}} H_n(y) .
\end{equation}

Then, we can define a new   learning function, denoted by $L_n: \mathcal{X} \mapsto \mathbb{R}$:
\begin{equation}
	L_n( \boldsymbol{x}| y^{\star}) = \sqrt{\varPhi \left( \frac{ y^{\star}-m_{g_n}\left( \boldsymbol{x} \right)}{\sigma _{g_n}\left( \boldsymbol{x} \right)} \right) \varPhi \left( -\frac{ y^{\star}-m_{g_n}\left( \boldsymbol{x} \right)}{\sigma _{g_n}\left( \boldsymbol{x} \right)} \right)}f_{\boldsymbol{X}}(\boldsymbol{x}).
\end{equation}
Note that this learning function is taken from the  integrand of the upper bound on the posterior standard deviation function of $F_Y$ and $\overline{F}_Y$ at $y=y^{\star}$, i.e., $\int_{\mathcal{X}} L_n( \boldsymbol{x}| y^{\star}) \mathrm{d} \boldsymbol{x}  = \overline{\sigma}_{F_{Y,n}}\left( y^{\star} \right) =\overline{\sigma}_{\overline{F}_{Y,n}}\left( y^{\star} \right)  $.  Therefore, it can be interpreted as a measure of the contribution at $\boldsymbol{x}$ to the total value of $\overline{\sigma}_{F_{Y,n}}\left( y^{\star} \right) $ or $\overline{\sigma}_{\overline{F}_{Y,n}}\left( y^{\star} \right) $.

Having defined the learning function, the next best point to evaluate the $g$ function can be selected by:
\begin{equation}\label{eq:next_best}
     \boldsymbol{x}^{(n+1)} =  \argmax_{\boldsymbol{x} \in \mathcal{X}} {L}_n ( \boldsymbol{x}).
\end{equation}

\subsection{Implementation procedure of the proposed method}

\begin{figure}[htb]
	\linespread{1}
	\centering
	\footnotesize
	\tikzstyle{startstop} = [rectangle, rounded corners, minimum width=1cm, minimum height=0.7cm,text centered, draw=red, fill=red!20]
	\tikzstyle{process} = [rectangle, minimum width=1cm, minimum height=0.7cm, text centered, draw=blue, align=center, fill=blue!20]
	\tikzstyle{decision} = [diamond, minimum width=1cm, minimum height=0.7cm, text centered, draw=green, fill=green!20, aspect=3]
	\tikzstyle{arrow} = [thin,->,>=stealth]
	
	\begin{tikzpicture}[node distance=1.3cm]
		\node (start) [startstop] {Start};
		
				\node (pro1) [process, below of=start] {Generate $N$ samples $\left\{ \boldsymbol{x}^{(j)}\right\}_{j=1}^N$ according to $f_{\boldsymbol{X}}(\boldsymbol{x})$ };   
		 
		\node (pro2) [process, below of=pro1] {Define an initial  dataset $\boldsymbol{\mathcal{D}}=\left\{ \boldsymbol{\mathcal{X}}, \boldsymbol{\mathcal{Y}}  \right\}$ and let $n=n_0$};   
		
		\node (pro3) [process, below of=pro2] {Obtain the posterior GP of the $g$-function conditional on $\boldsymbol{\mathcal{D}}$};
		
		\node (pro4) [process, below of=pro3, yshift=-0.30 cm] {Calculate  $\hat{m}_{F_{Y,n}}\left( y_{t} \right)$,  $\hat{m}_{\overline{F}_{Y,n}}\left( y_{t} \right)$ and $ \hat{\overline{\sigma}}_{F_{Y,n}}\left( y_{t} \right) $ ($ \hat{\overline{\sigma}}_{\overline{F}_{Y,n}}\left( y_{t} \right) $); \\ Obtain $\hat{\overline{\mathrm{CoV}}} _{F_{Y,n}} (y_t) $ and $\hat{\overline{\mathrm{CoV}}} _{\overline{F}_{Y,n}}  (y_t) $};

		\node (pro5) [decision,below of=pro4,yshift=-0.6 cm] {Stopping criterion?};

		\node (pro6) [process,below of=pro4, yshift=-0.6 cm, xshift=7.5 cm] {Identify  $\boldsymbol{x}^{(n+1)}$  by maximizing the $L_n$ function; \\ Evaluate the corresponding $g$-fucntion value  $y^{(n+1)}$; \\ Enrich the previous dataset   $\boldsymbol{\mathcal{D}}$ with $\left\{\boldsymbol{u}^{(n+1)}, y^{(n+1)} \right\}$. \\ Let $n=n+1$};
		
		\node (pro7) [process,below of=pro5, yshift=-0.6 cm] {Calculate $\hat{m}_{f_{Y,n}}(y_t)$;  \\ Return $\hat{m}_{F_{Y,n}}\left( y_{t} \right)$,  $\hat{m}_{\overline{F}_{Y,n}}\left( y_{t} \right)$,  $\hat{\overline{\mathrm{CoV}}} _{F_{Y,n}} (y_t)$,  $\hat{\overline{\mathrm{CoV}}} _{\overline{F}_{Y,n}}  (y_t)$ and $\hat{f}_{Y,n}(y_t)$};
		
		\node (stop) [startstop, below of=pro7, yshift=-0.25 cm] {Stop};	
		
		\draw [arrow] (start) -- (pro1);
		\draw [arrow] (pro1) -- (pro2);
		\draw [arrow] (pro2) -- (pro3);
		\draw [arrow] (pro3) -- (pro4);
		\draw [arrow] (pro4) -- (pro5);
		\draw [arrow] (pro5) -- node[above]  {No} (pro6);
		\draw [arrow] (pro6) |-  (pro3);
		\draw [arrow] (pro5) -- node[right]  {Yes} (pro7);
		\draw [arrow] (pro7) -- (stop);	
	\end{tikzpicture}
	\caption{Flowchart of the proposed BAL method.}
	\label{fig:flowchart}
\end{figure}
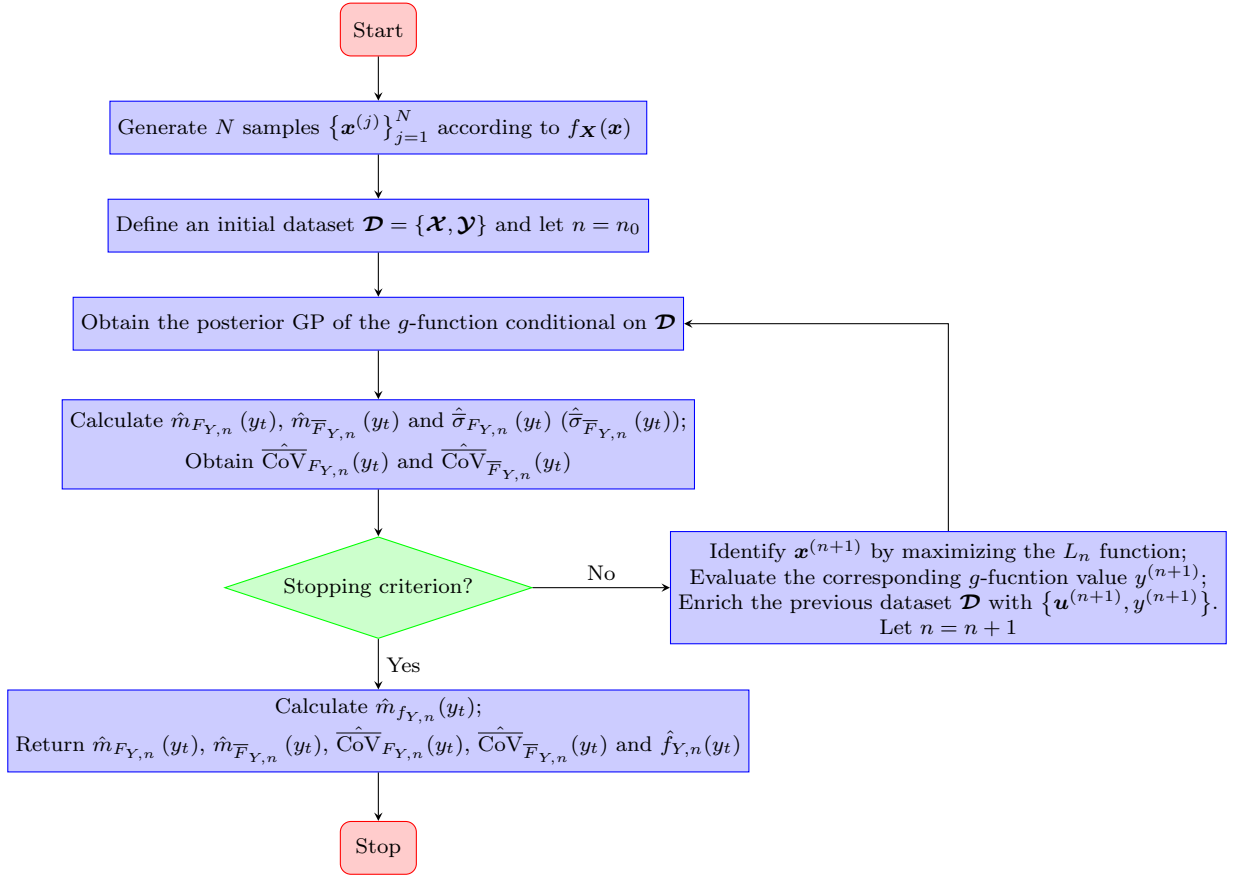

The implementation procedure of the proposed Bayesian active learning method can be summarized in seven main steps, accompanied by a flowchart in Fig. \ref{fig:flowchart}.

\textbf{Step 1: Generate  $N$ samples according to $f_{\boldsymbol{X}}(\boldsymbol{x})$}  

Since several intractable integrals (i.e., $m_{F_{Y,n}}\left( y \right)$,  $m_{\overline{F}_{Y,n}}\left( y \right)$,  $m_{f_{Y,n}}\left( y \right)$,  $ \overline{\sigma}_{F_{Y,n}}\left( y \right) $ ($ \overline{\sigma}_{\overline{F}_{Y,n}}\left( y \right) $)) entail numerical integration under the same density $f_{\boldsymbol{X}}(\boldsymbol{x})$, we first generate a set of $N$ samples according to $f_{\boldsymbol{X}}(\boldsymbol{x})$ using a suitable low-discrepancy sequence (Sobol sequence in this study), which are denoted as $\left\{ \boldsymbol{x}^{(j)}\right\}_{j=1}^N$.  The number of samples $N$ depends on the  expected statistical error for approximating these integrals.

\textbf{Step 2: Define  an initial observation dataset}  

The Bayesian active learning process needs to be initialized with an initial set of observations. First, a small number (say $n_0$) of uniform points, $\boldsymbol{\mathcal{X}} = \left\{ \boldsymbol{x}^{(i)}\right\}_{i=1}^{n_0}$, are generated within a hyper-rectangular $ \varLambda_1 = \prod_{r=1}^{d}  \left[ a_r, b_r \right] \subseteq \mathcal{X}$ using an appropriate low-discrepancy sequence (Hammersley sequence in this study).  In this study,  the  lower and upper bounds in the  $r$-th dimension  are specified by: $a_r = F_{X_r}^{-1} (\rho)$  and $b_r = F_{X_r}^{-1} (1-\rho_1)$, where $F_{X_r}$ denotes the marginal CDF of $X_r$ and $\rho_1$ is  a small truncation probability.  Second, the $g$-function is evaluated at $\boldsymbol{\mathcal{X}}$ to produce the corresponding response values, i.e., $\boldsymbol{\mathcal{Y}} = \left\{y^{(i)}\right\}_{i=1}^{n_0}$ with $y^{(i)} = g (\boldsymbol{x}^{(i)})$. Finally, the initial dataset is formed by $\boldsymbol{\mathcal{D}} = \left\{  \boldsymbol{\mathcal{X}}, \boldsymbol{\mathcal{Y}}  \right\} $. Let $n=n_0$.

\textbf{Step 3:  Obtain the posterior statistics of $g$}

This step involves obtaining the  posterior GP  of $g$ conditional on $\boldsymbol{\mathcal{D}} $, i.e., $g_n (\boldsymbol{x}) \sim   \mathcal{GP}(m_{g_n} (\boldsymbol{x}), k_{g_n} (\boldsymbol{x}, \boldsymbol{x}^{\prime}) )$.  Such a task can be performed by many well-developed GP regression toolboxes. In this study, we employ the \emph{fitrgp} function available in  the Statistics and Machine Learning Toolbox of Matlab R2024a.

\textbf{Step 4:  Evaluate the posterior statistics of $F_Y$ and $\overline{F}_Y$}

\indent    \indent  \textbf{Step 4.1} Evaluate the posterior mean and standard deviation functions of $g$ at $\left\{ \boldsymbol{x}^{(j)}\right\}_{j=1}^N$, i.e.,  $\boldsymbol{M}=  \left\{   m_{g_n} (\boldsymbol{x}^{(j)}) \right\}_{j=1}^N $ and $\boldsymbol{	\varXi}= \left\{ \sigma _{g_n}\left( \boldsymbol{x}^{(j)} \right)  \right\} _{j=1}^N$;

\indent    \indent  \textbf{Step 4.2} Discretize  the range of interest $\left[y_{\mathrm{min}},  y_{\mathrm{max}}\right]$ into $h$ equally spaced sub-intervals, i.e.,  $y_{\mathrm{min}}=y_0<y_1<\cdots <y_{h} = y_{\mathrm{max}}$ with spacing $(y_{\mathrm{max}} - y_{\mathrm{min}})/h$. Note that the lower and upper bounds $y_{\mathrm{min}}$ and $y_{\mathrm{max}}$ are difficult to know a priori.  Therefore, we propose to estimate them from the posterior GP of $g$.  More specifically,   $y_{\mathrm{min}}$ and $y_{\mathrm{max}}$ are specified by the $p$ and $1-p$ quantiles of $\boldsymbol{M} - 	\lambda \boldsymbol{	\varXi}$ and $\boldsymbol{M} + \lambda \boldsymbol{	\varXi}$, respectively,  where $\lambda$ is introduced to account for the posterior standard deviation. In this manner, the region of interest of the response CDF and CCDF is expected to be that with probability within $p$ and $1-p$,  where the value of $p$ can be specified according to the requirements of practical applications.

\indent    \indent  \textbf{Step 4.3} First, obtain the estimates $\hat{m}_{F_{Y,n}}\left( y_{t} \right)$,  $\hat{m}_{\overline{F}_{Y,n}}\left( y_{t} \right)$ and $ \hat{\overline{\sigma}}_{F_{Y,n}}\left( y_{t} \right) $ ($ \hat{\overline{\sigma}}_{\overline{F}_{Y,n}}\left( y_{t} \right) $) by using $\boldsymbol{M}$ and $\boldsymbol{	\varXi}$, $t=0,1, \cdots, h$. Then, calculate the upper bound estimates on the posterior CoV functions of the response CDF and CCDF by $\hat{\overline{\mathrm{CoV}}} _{F_{Y,n}} (y_t) =  \frac{\hat{\overline{\sigma}}_{F_{Y,n}}\left( y_t \right) }{\hat{m}_{F_{Y,n}}\left( y_t \right)}$ and $\hat{\overline{\mathrm{CoV}}} _{\overline{F}_{Y,n}}  (y_t) = \frac{\hat{\overline{\sigma}}_{\overline{F}_{Y,n}}\left( y_t \right)}{\hat{m}_{\overline{F}_{Y,n}}\left( y_t \right)}$.

\textbf{Step 5:  Check  the  stopping criterion}

If the stopping criterion $\max_{y_t} \hat{H}_n(y_t)  < \epsilon$ is satisfied twice in a row, where $\hat{H}_n(y_t) = \max \left(   \hat{\overline{\mathrm{CoV}}} _{F_{Y,n}} (y_t) ,  \hat{\overline{\mathrm{CoV}}} _{\overline{F}_{Y,n}}  (y_t)  \right)$, then  go to \textbf{Step 7}; otherwise, go to \textbf{Step 6}.

\textbf{Step 6:    Enrich the observation dataset}

In this stage, the previous observation dataset needs to be enriched. First, identify the critical location $y^{\star} = \argmax_{y_t } \hat{H}_n(y_t)$ and identify the next best point $\boldsymbol{x}^{(n+1)}$ by  $\boldsymbol{x}^{(n+1)} =  \argmax_{\boldsymbol{x} \in \varLambda_2} {L}_n ( \boldsymbol{x})$ (Genetic algorithm is used in this study), $\varLambda_2$ is specified in similar with   $\varLambda_1$  by replacing $\rho_1$ with $\rho_2$. Then, the $g$-function is evaluated at $\boldsymbol{x}^{(n+1)}$ to obtain the corresponding output $y^{(n+1)}$, i.e., $y^{(n+1)} = g (\boldsymbol{x}^{(n+1)}) $. At last, the previous dataset $\boldsymbol{\mathcal{D}}$ is enriched with $\left\{ \boldsymbol{x}^{(n+1)}, y^{(n+1)} \right\}$, i.e., $\boldsymbol{\mathcal{D}} = \boldsymbol{\mathcal{D}} \cup \left\{ \boldsymbol{x}^{(n+1)}, y^{(n+1)} \right\}$. Let $n=n+1$ and go to \textbf{Step 3}.

\textbf{Step 7: Evaluate the posterior mean of  $f_{Y}$ and return the results}

Calculate $\hat{m}_{f_{Y,n}}(y_t)$ using the current $\boldsymbol{M}$ and $\boldsymbol{	\varXi}$, $t=0,1,\cdots,h$. Return $\hat{m}_{F_{Y,n}}\left( y_{t} \right)$,  $\hat{m}_{\overline{F}_{Y,n}}\left( y_{t} \right)$,  $\hat{\overline{\mathrm{CoV}}} _{F_{Y,n}} (y_t)$,  $\hat{\overline{\mathrm{CoV}}} _{\overline{F}_{Y,n}}  (y_t)$ and $\hat{f}_{Y,n}(y_t)$ as the final results.

\section{Numerical examples}\label{sec:examples}

In this section, five numerical examples are investigated to demonstrate the performance of the proposed  BAL method for estimating the response CDF, CCDF and PDF.  The parameters of our method is set as  follows: $N = 5 \times 10^5$, $n_0  = 10$, $\rho_1 = 10^{-5}$, $\rho_2 = 10^{-8}$, $h=100$, $p= 5 \times 10^{-5}$, $\lambda=2$ and  $\epsilon=0.20$. For comparison, one of the representative existing methods, called  active learning- based GP (AL-GP) metamodelling  method \cite{wang2020novel},  is also conducted for the first three examples.  The number of training candidates is set to $5 \times 10^5$ with other parameters specified in each example. For both the proposed BAL method and the AL-GP method, 20 independent runs are performed to assess their robustness. When no analytical (or semi-analytical) solution is available, MCS is used to produce a reference result.

\subsection{Example 1: A toy example}
The first example considers a toy example taken from \cite{wang2020novel}:
\begin{equation}
	Y = g(\boldsymbol{X}) = \min \left[ X_1 - X_2, X_1 + X_2 \right], 
\end{equation}
where $X_1$ and $X_2$ are two independent standard normal  variables.  The CDF, CCDF and PDF of $Y$ can be expressed as:
\begin{equation}\label{eq:CDF_toy_exam}
     F_{Y} (y) = \varPhi \left(\frac{y}{\sqrt{2}}\right) \left(2-  \varPhi \left(\frac{y}{\sqrt{2}}\right)  \right),
\end{equation}
\begin{equation}
    \overline{F}_{Y} (y) = 1- \varPhi \left(\frac{y}{\sqrt{2}}\right) \left(2-  \varPhi \left(\frac{y}{\sqrt{2}}\right)  \right),
\end{equation}
\begin{equation}\label{eq:pDF_toy_exam}
	f_{Y} (y) = \sqrt{2}  \phi\left(\frac{y}{\sqrt{2}}\right)  \left(1 - \varPhi\left(\frac{y}{\sqrt{2}}\right)\right).
\end{equation}
Although $\varPhi$ does not have an analytical expression in terms of elementary functions, it  can be computed accurately using numerical methods or approximations. Therefore, we still consider the results of Eqs. \eqref{eq:CDF_toy_exam}  - \eqref{eq:pDF_toy_exam} as `exact'.

\begin{figure}[htb]
	\centering
	\subfigure[CDF]{
		\begin{minipage}{8.0cm}
			\centering
			\includegraphics[scale=0.35]{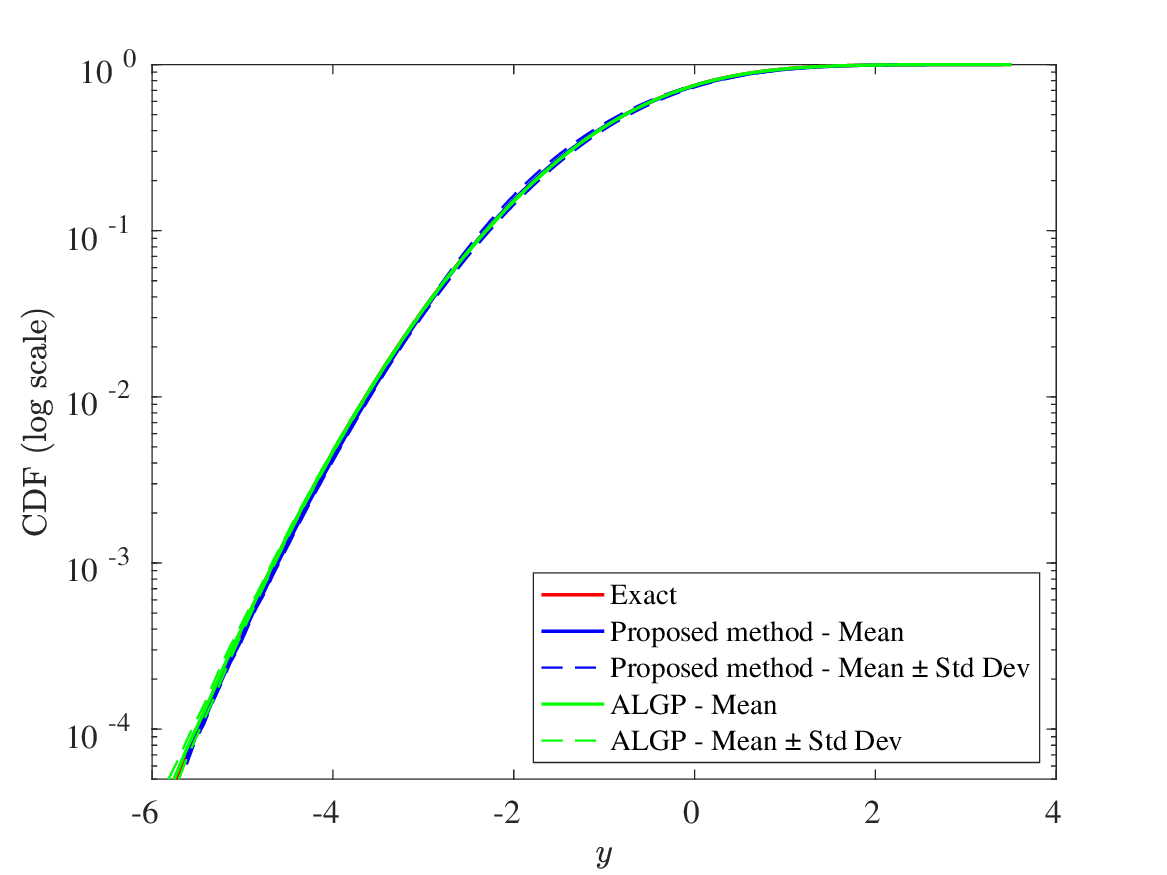}\label{fig:cdf_exam_1_a}
		\end{minipage}	
	}%
	\subfigure[CCDF]{
		\begin{minipage}{8.0cm}
			\centering
			\includegraphics[scale=0.35]{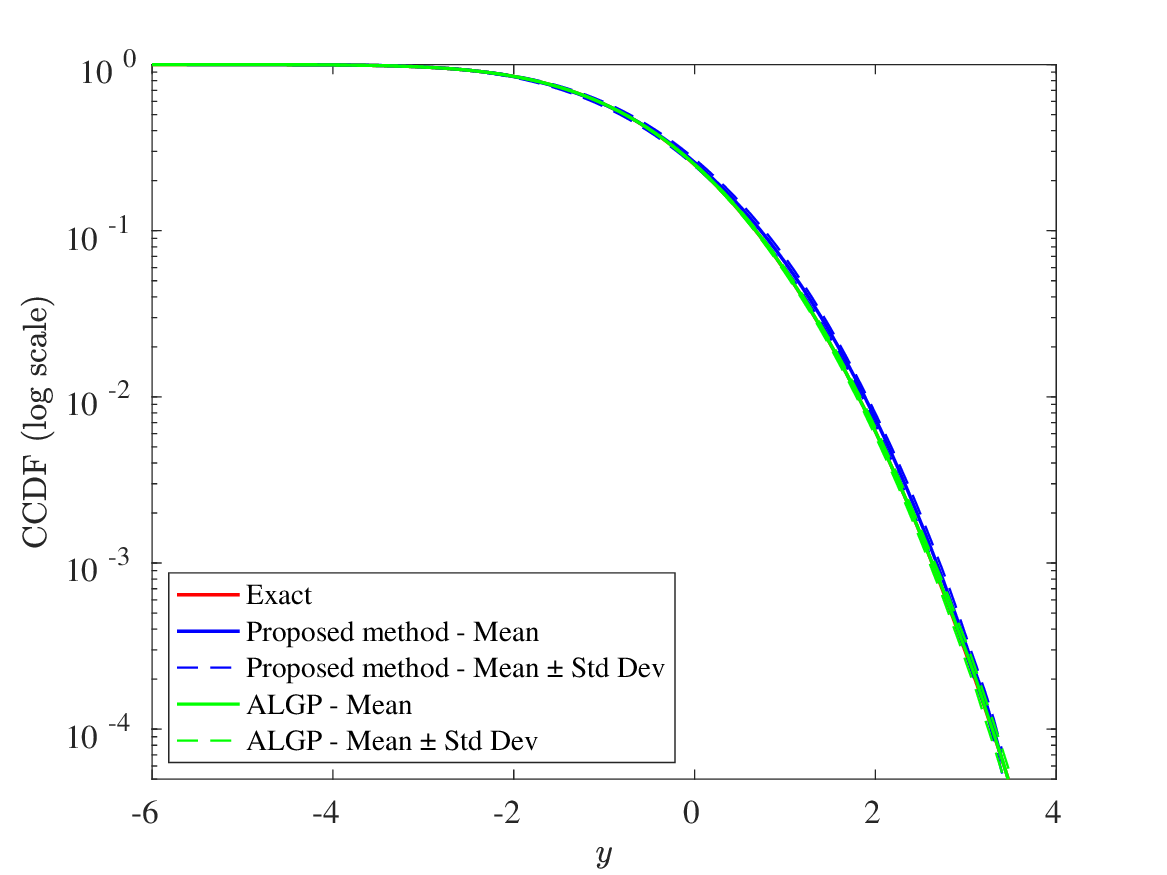}\label{fig:cdf_exam_1_b}
		\end{minipage}	
	}%

	\subfigure[Upper bound on the posterior CoV function of CDF]{
	\begin{minipage}{8.0cm}
		\centering
		\includegraphics[scale=0.35]{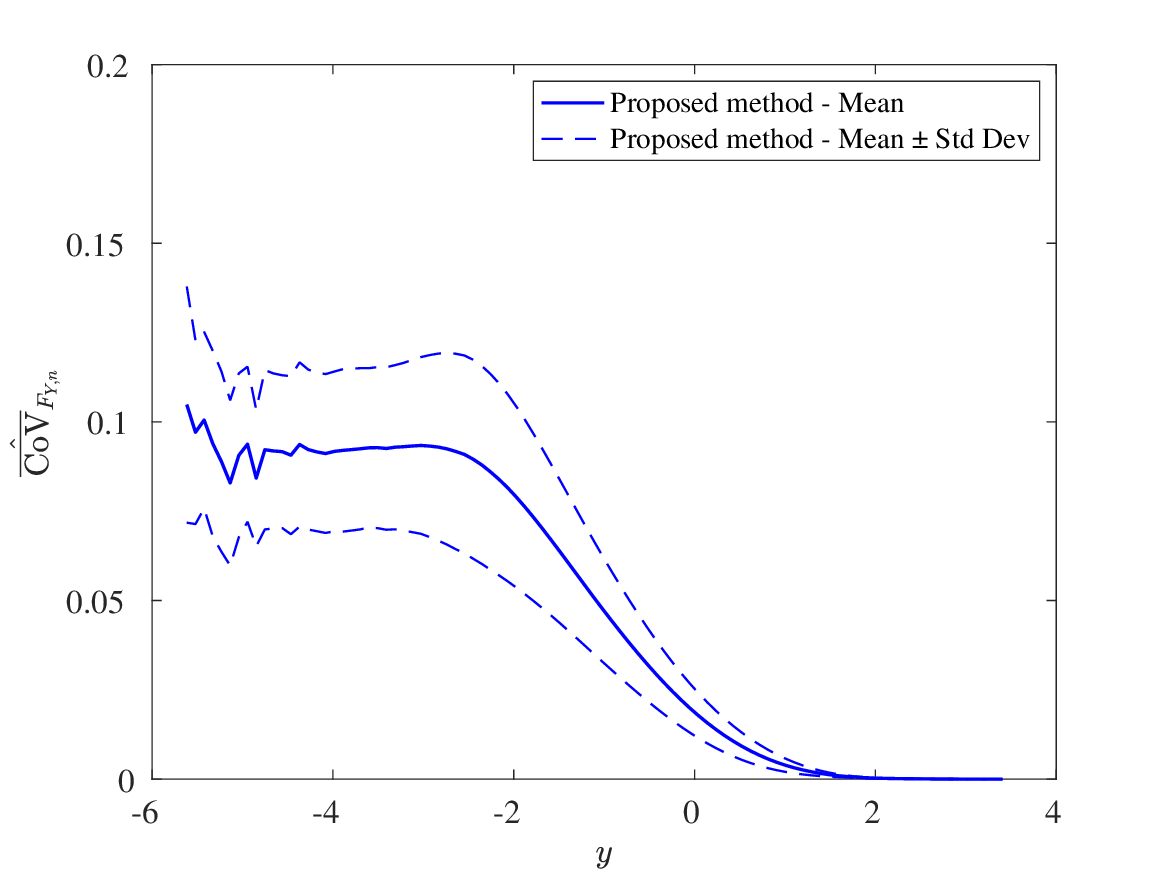}\label{fig:cdf_exam_1_c}
	\end{minipage}	
}%
\subfigure[Upper bound on the posterior CoV function of CCDF]{
	\begin{minipage}{8.0cm}
		\centering
		\includegraphics[scale=0.35]{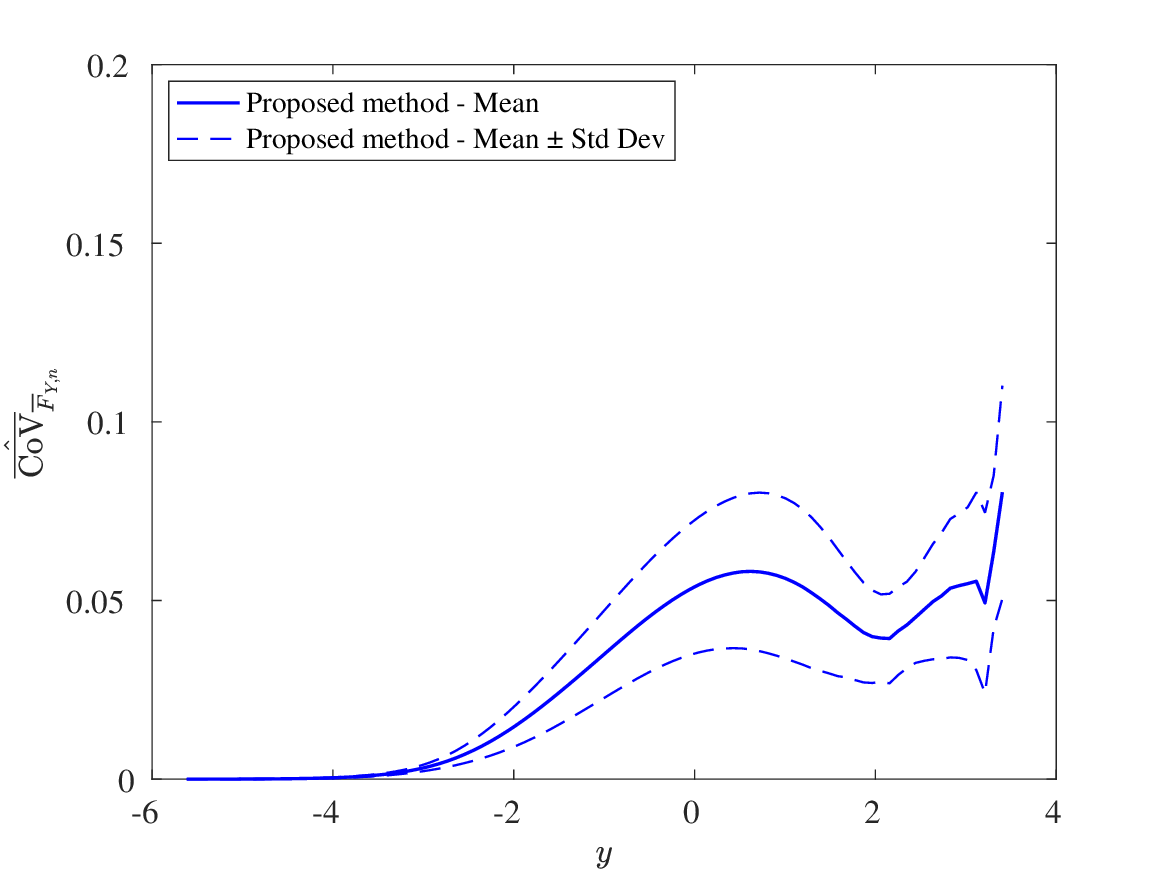}\label{fig:cdf_exam_1_d}
	\end{minipage}	
}%
	
	\caption{Response CDF and CCDF  for Example 1.}
	\label{fig:CDF_exam_1}
\end{figure}

\begin{figure}[htb]
	\centering
     \includegraphics[scale=0.35]{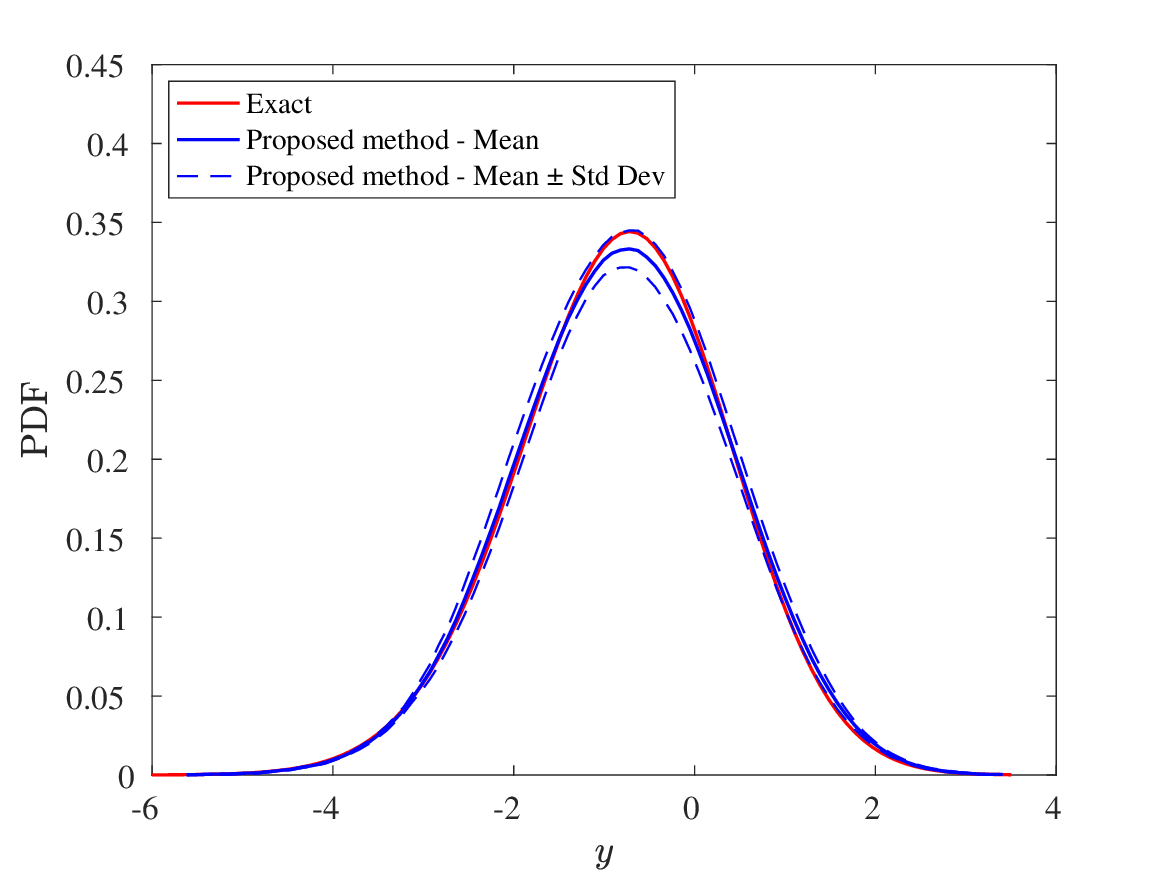}
     \caption{Response PDF  for Example 1.}\label{fig:PDF_exam_1}
\end{figure}

For the AL-GP method, the range of interest is set to $[-6.0,3.5]$ and the stopping criterion threshold is set to $\overline{\epsilon} = 0.15$.  Fig.  \ref{fig:CDF_exam_1} shows the results of the response CDF and CCDF. As can be observed in  Figs. \ref{fig:cdf_exam_1_a} and \ref{fig:cdf_exam_1_b},  both the AL-GP method and the proposed BAL method produce  CDF/CCDF mean curves that are close to the exact ones, along with notably narrow mean ± std dev bands.  It is worth noting that  the proposed method can provide the error measures for the response CDF/CCDF, i.e., the upper bounds of the CoV functions of the response CDF and CCDF, as depicted in \ref{fig:cdf_exam_1_c} and \ref{fig:cdf_exam_1_d}, respectively.  It can be seen that the mean curves and also the mean $+$ std dev curves are all well below the $\epsilon = 0.20$ threshold.  Additionally, the proposed method can also produce the response PDF as a by-product, with statistical results shown in Fig. \ref{fig:PDF_exam_1}. The PDF mean curve is close to the exact one and also features a narrow mean ± std dev band. On average, the proposed method requires significantly fewer $g$-function calls than the AL-GP method. Additionally, the AL-GP method exhibits a higher CoV for the number of $g$-function  evaluations compared to the proposed method.

\begin{table}[htb]
	\centering
	\caption{Comparison of the number of $g$-function calls for Example 1.}\label{tab:Ncall_exam_1}
	\begin{tabular}{lll}
		\hline
 \multirow{2}{*}{Method}		&  \multicolumn{2}{c}{$N_{\mathrm{call}}$}  \\ \cline{2-3}
 
   	                                                 &  Mean  &  CoV  \\
	\hline
	Proposed method &  10 + 15.00  = 25.00   & 12.91\%  \\
	AL-GP    &  12 + 33.20 =  45.20 &  50.74\%    \\
	\hline
	\end{tabular}
\end{table}

\begin{figure}[htb]
  \centering
  \includegraphics[scale=0.40]{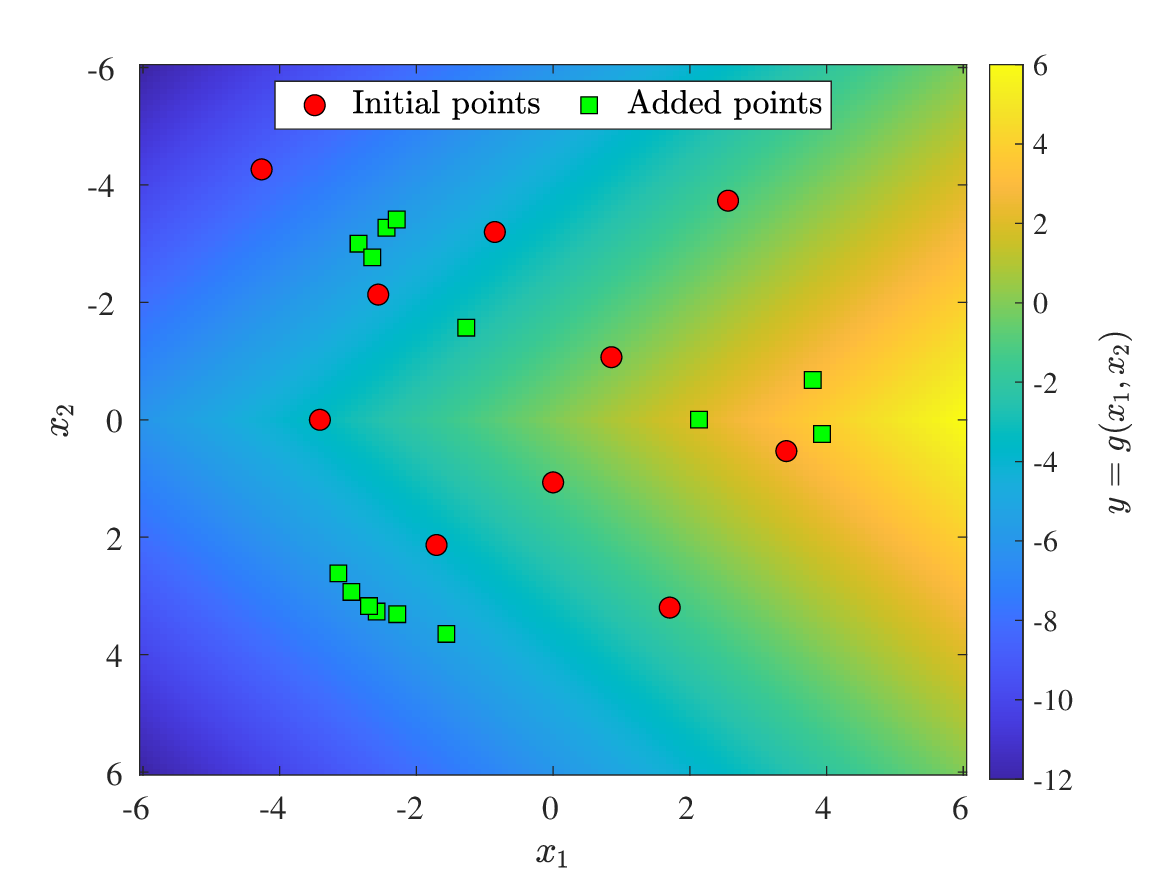}
  \caption{Design of computer experiments for Example 1. }\label{fig:points_exam_1}
\end{figure}

For illustrative purposes, the points selected during an arbitrary run of our approach are shown in Fig. \ref{fig:points_exam_1}.  As can be seen, the 10 initial points are evenly distributed in the input space, as expected.  During the active learning phase, only 14 additional points are identified  before the proposed stopping criterion is met, which strategically chosen by maximizing the  proposed learning function.

\subsection{Example 2: The Ishigami function}

As a second example, consider the Ishigami function:
\begin{equation}
	Y = \sin(X_1) + a \sin^2(X_2) + b X_3^4\sin(X_1),
\end{equation}
where $a=7$ and $b=0.1$; $X_1$, $X_2$ and $X_3$ are three independent uniform random variables within $\left[-\pi,\pi\right]$.

\begin{figure}[htb]
	\centering
	\subfigure[CDF]{
		\begin{minipage}{8.0cm}
			\centering
			\includegraphics[scale=0.35]{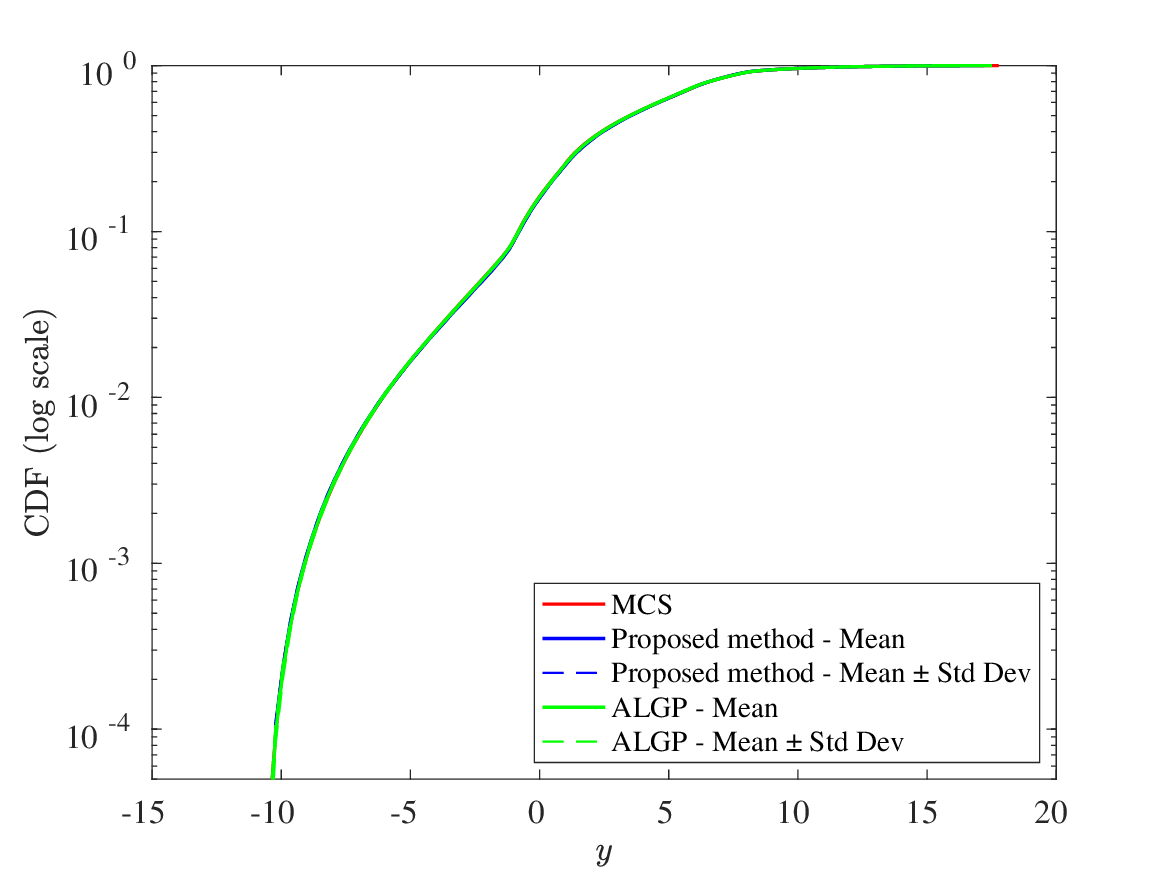}\label{fig:cdf_exam_2_a}
		\end{minipage}	
	}%
	\subfigure[CCDF]{
		\begin{minipage}{8.0cm}
			\centering
			\includegraphics[scale=0.35]{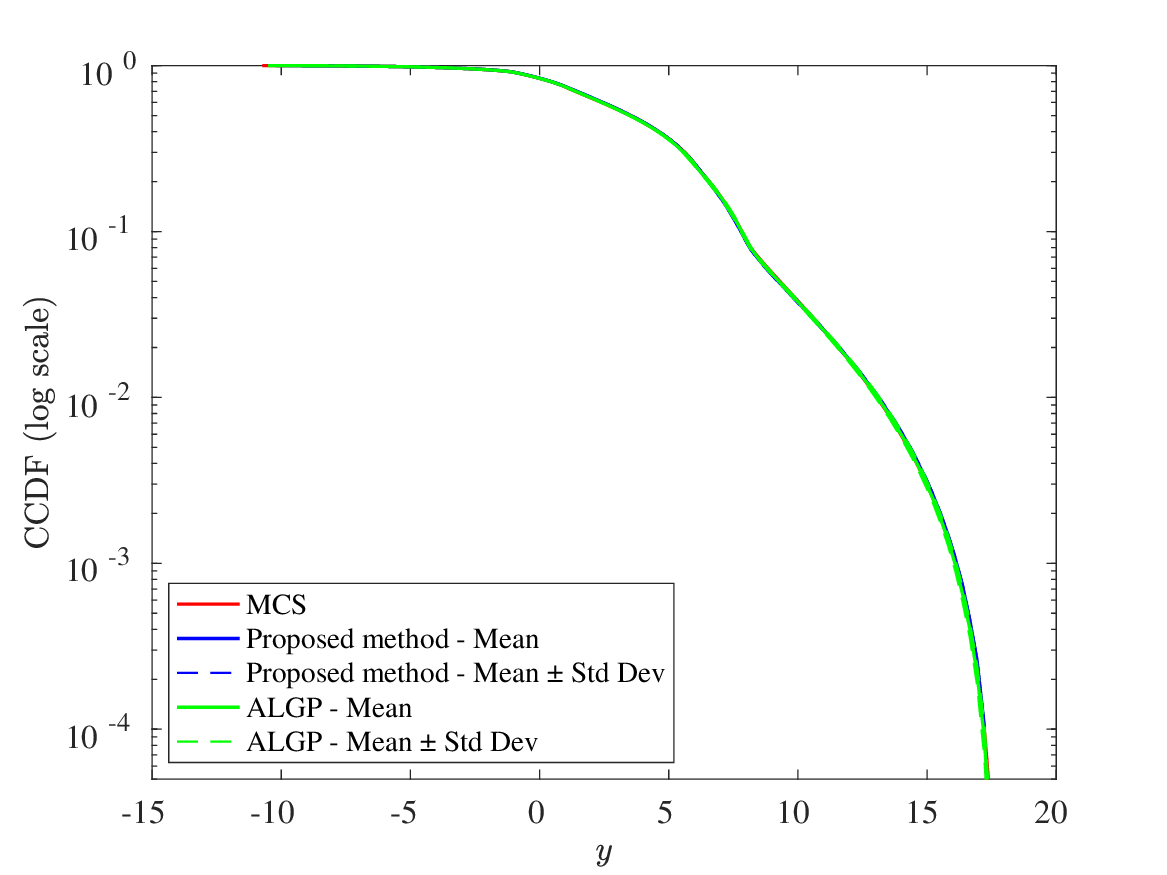}\label{fig:cdf_exam_2_b}
		\end{minipage}	
	}%
	
	\subfigure[Upper bound on the posterior CoV function of CDF]{
		\begin{minipage}{8.0cm}
			\centering
			\includegraphics[scale=0.35]{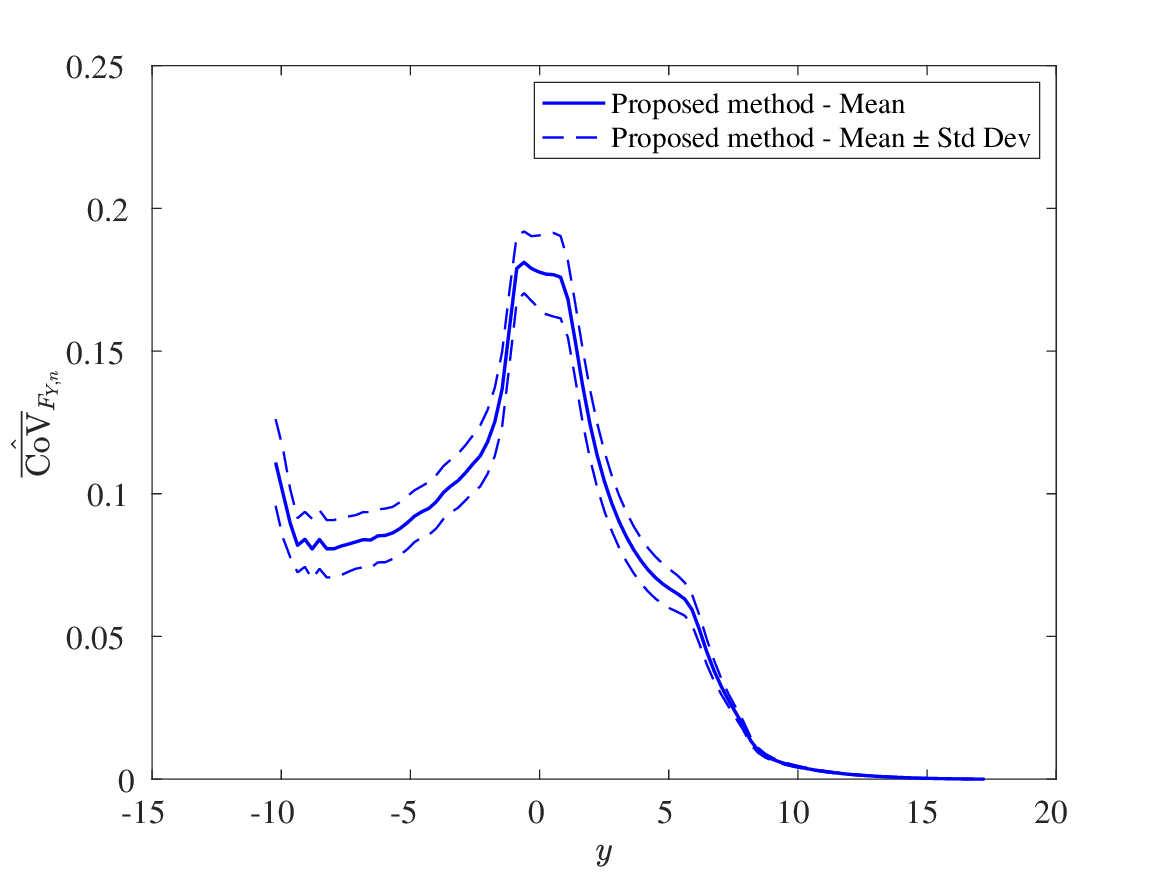}\label{fig:cdf_exam_2_c}
		\end{minipage}	
	}%
	\subfigure[Upper bound on the posterior CoV function of CCDF]{
		\begin{minipage}{8.0cm}
			\centering
			\includegraphics[scale=0.35]{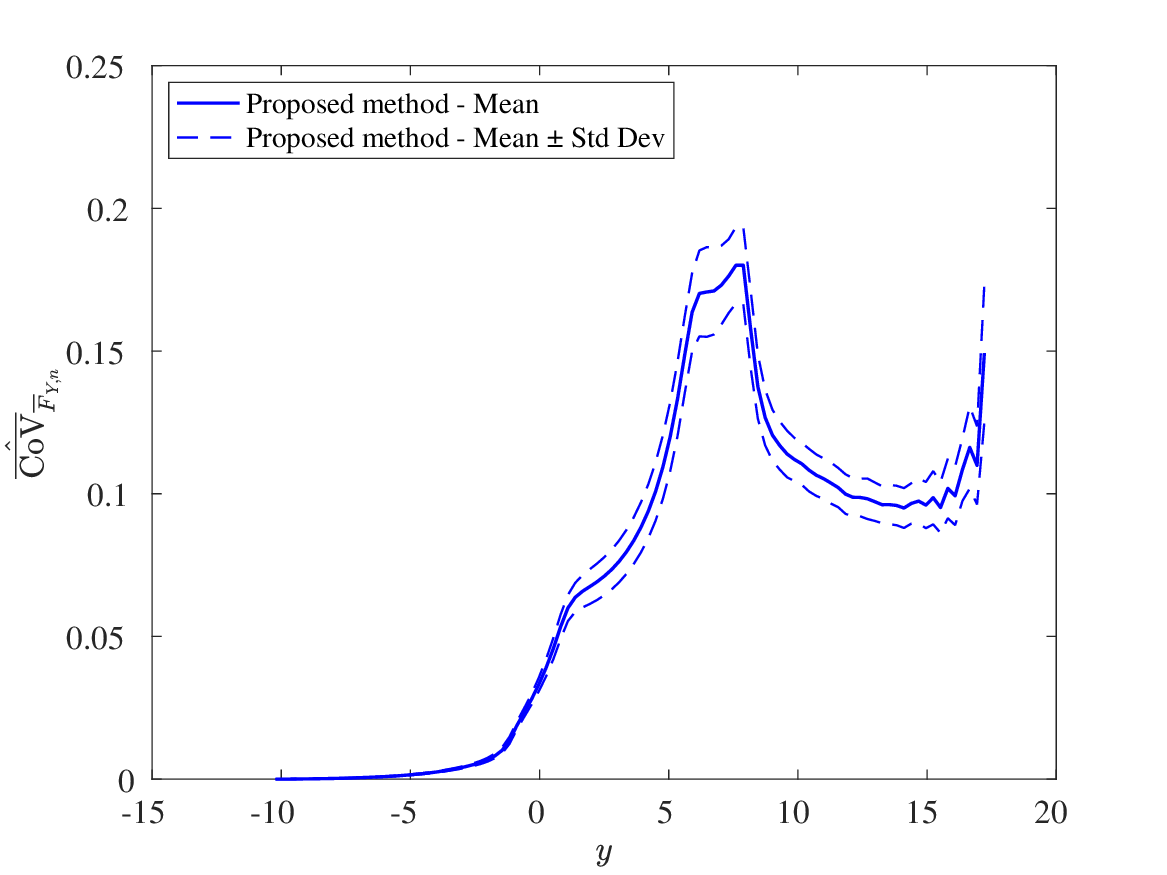}\label{fig:cdf_exam_2_d}
		\end{minipage}	
	}%
	
	\caption{Response CDF and CCDF  for Example 2.}
	\label{fig:CDF_exam_2}
\end{figure}

\begin{figure}[htb]
	\centering
	\includegraphics[scale=0.35]{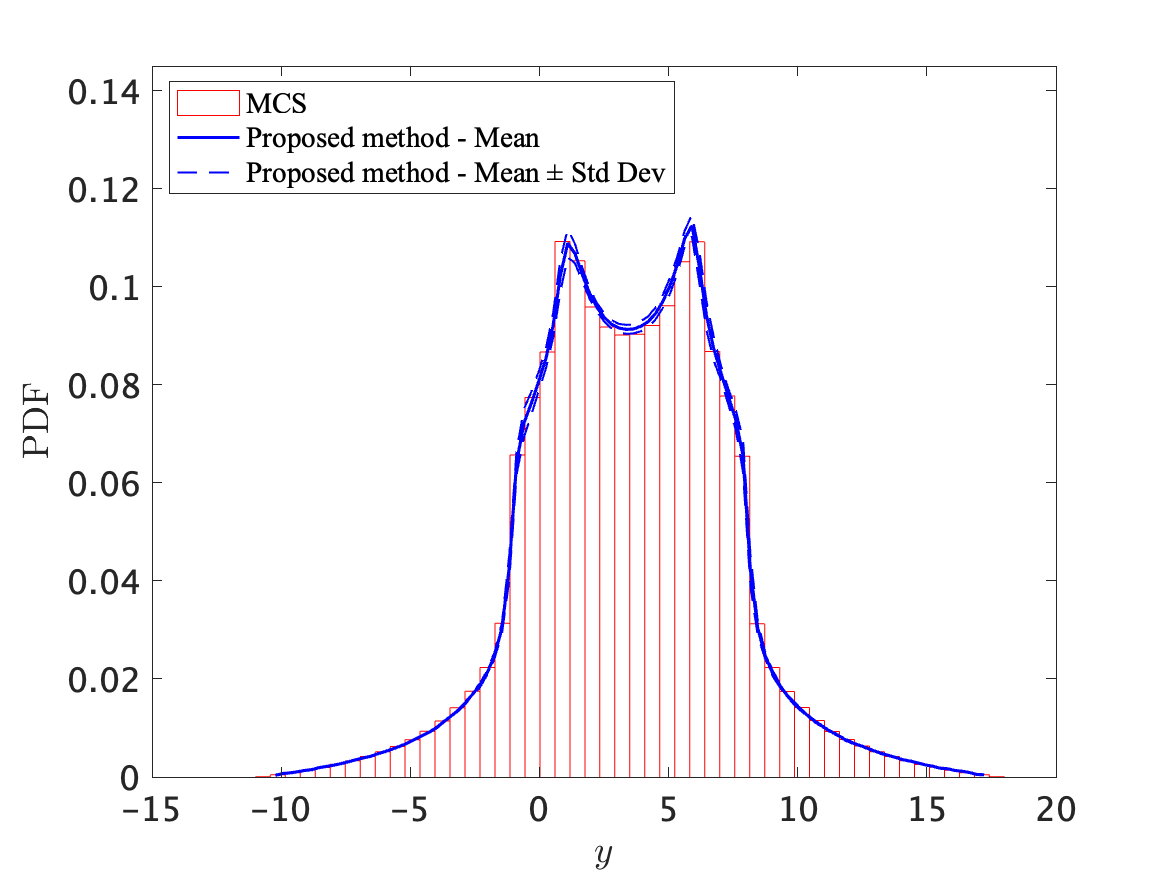}
	\caption{Response PDF  for Example 2.}\label{fig:PDF_exam_2}
\end{figure}

The reference solutions of the response CDF, CCDF and PDF are generated using  MCS with $10^7$ samples.  The range of interest and stopping criterion threshold of the AL-GP method are are set to $[-10.5,17.5]$  and $\overline{\epsilon} = 0.15$, respectively.  The results for the response CDF and CCDF are shown in Fig.  \ref{fig:CDF_exam_2}.  Figs. \ref{fig:cdf_exam_2_a} and \ref{fig:cdf_exam_2_b} indicate that both the AL-GP method and the proposed method can yield highly accurate response CDF and CCDF, even in the low-probability range.   The proposed method is  capable of providing  the upper bounds of the posterior CoV functions of the response CDF and CCDF, as  presented in  Figs.  \ref{fig:cdf_exam_2_c} and  \ref{fig:cdf_exam_2_d}  respectively. As can be observed, the mean curves are situated behind the threshold value $\epsilon=0.20$, and the mean ± standard deviation bands are notably narrow.  Fig. \ref{fig:PDF_exam_2} depicts the by-product PDF generated by the proposed method, which is compared to the histogram produced by MCS. It is noteworthy that the proposed method is able to accurately capture the bi-modal shape.  Table \ref{tab:Ncall_exam_2} indicates that the proposed method necessitates fewer $g$-function evaluations on average and exhibits a slightly smaller  CoV than the AL-GP method.

\begin{table}[htb]
	\centering
	\caption{Comparison of the number of $g$-function calls for Example 2.} \label{tab:Ncall_exam_2}
	\begin{tabular}{lll}
		\hline
		\multirow{2}{*}{Method}		&  \multicolumn{2}{c}{$N_{\mathrm{call}}$}  \\ \cline{2-3}
		
		&  Mean  &  CoV  \\
		\hline
		Proposed method &  10 + 187.00  = 197.00   & 2.88\%  \\
		AL-GP    &  12 + 226.55 =  238.55 &  3.93\%    \\
		\hline
	\end{tabular}
\end{table}

\subsection{Example 3: A nonlinear oscillator}

\begin{figure}[htb]
	\centering
	\includegraphics[scale=0.30]{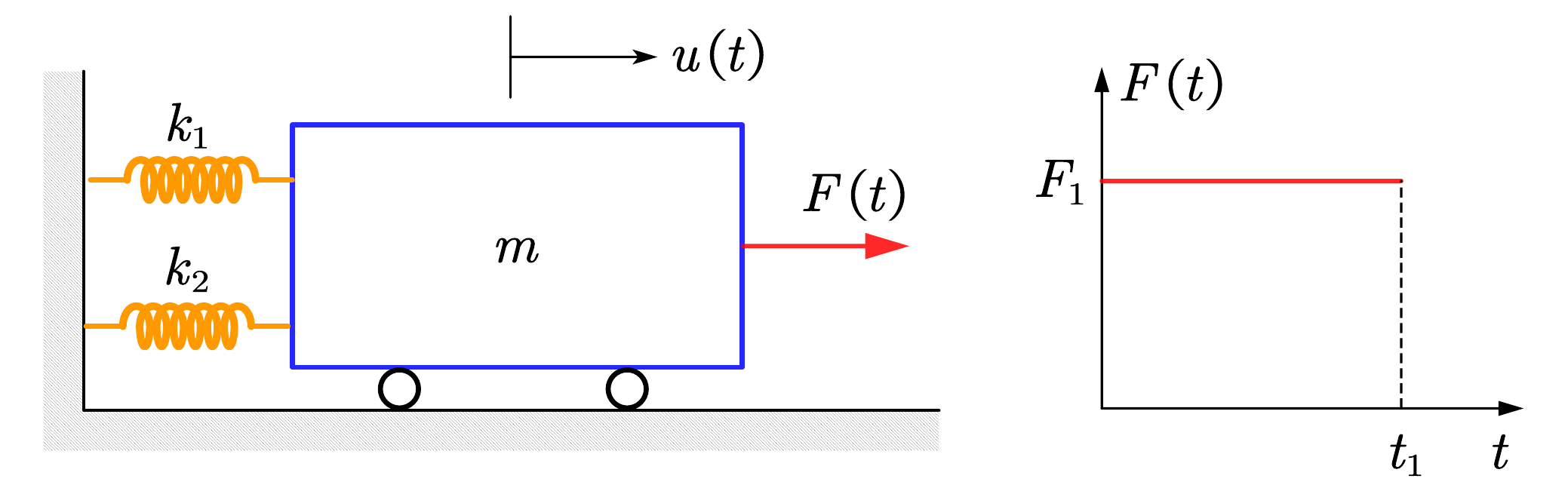}
	\caption{A nonlinear oscillator under a  rectangular pulse load.}\label{fig:oscillator}
\end{figure}

The third numerical example involves a single-degree-of-freedom nonlinear oscillator under a  rectangular pulse load \cite{bucher1990fast}, as shown in Fig. \ref{fig:oscillator}. The response function is defined in terms of the performance function of the oscillator:
\begin{equation}
Y =	g(\boldsymbol{X})=3 r-\left|\frac{2 F_1}{k_1+k_2} \sin \left(\frac{t_1}{2} \sqrt{\frac{k_1+k_2}{m}}\right)\right|,
\end{equation}
where $m$, $k_1$, $k_2$, $r$, $F_1$ and $t_1$ are six independent random variables, as listed in Table \ref{tab:rv_exam_3}.

\begin{table}[htb]
    \centering
    \caption{Input random variables for  Example 3.}\label{tab:rv_exam_3}
    \begin{tabular}{llll}
         \hline
      Variable    &  Distribution  &  Mean  & Std Dev  \\
           \hline
            $m$             &  Normal          &         1.0     &       0.05           \\
            $k_1$             &  Normal          &       1.0        &        0.10         \\
            $k_2$             &  Normal          &       0.2        &         0.01         \\
             $r$             &  Normal          &            0.5   &             0.05     \\
             $F_1$             &  Normal          &         1.0      &             0.20     \\
             $t_1$             &  Normal          &            1.0   &                0.20  \\
            \hline
    \end{tabular}
\end{table}

\begin{figure}[htb]
	\centering
	\subfigure[CDF]{
		\begin{minipage}{8.0cm}
			\centering
			\includegraphics[scale=0.35]{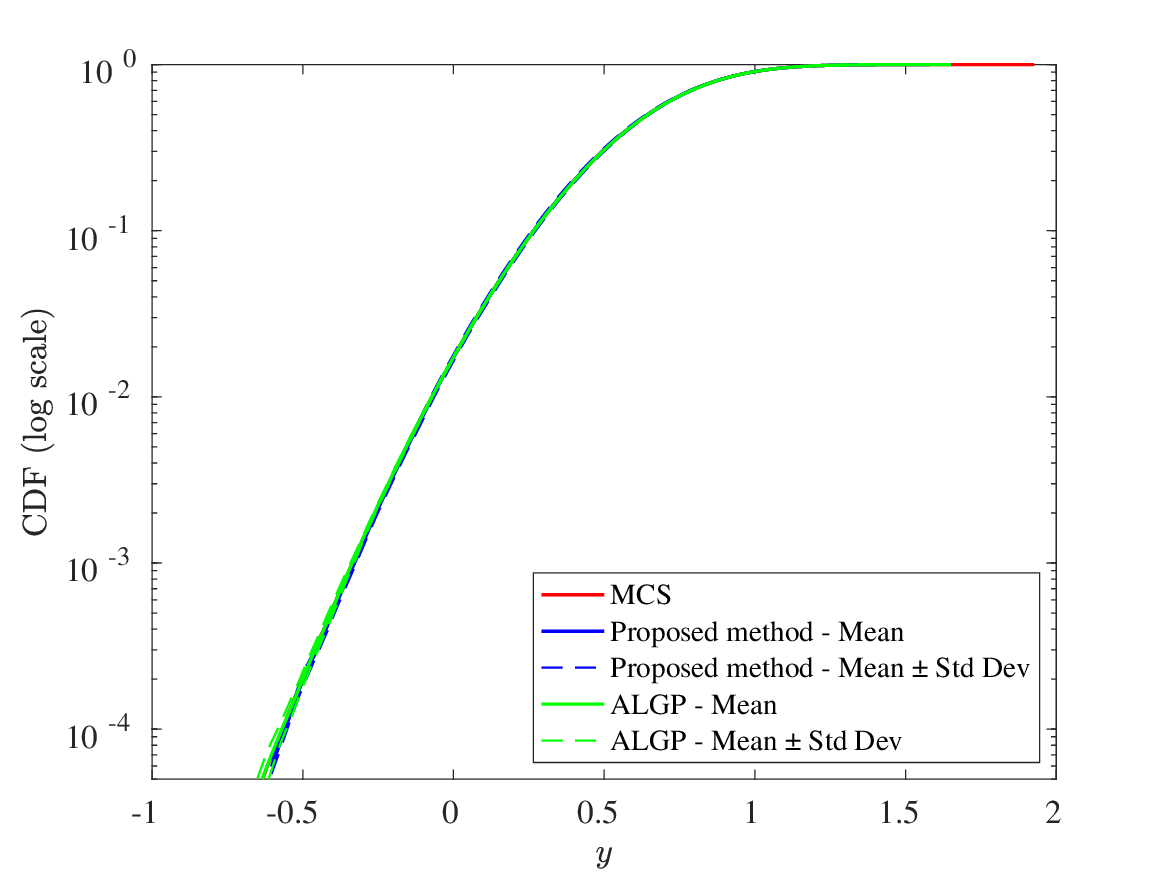}\label{fig:cdf_exam_3_a}
		\end{minipage}	
	}%
	\subfigure[CCDF]{
		\begin{minipage}{8.0cm}
			\centering
			\includegraphics[scale=0.35]{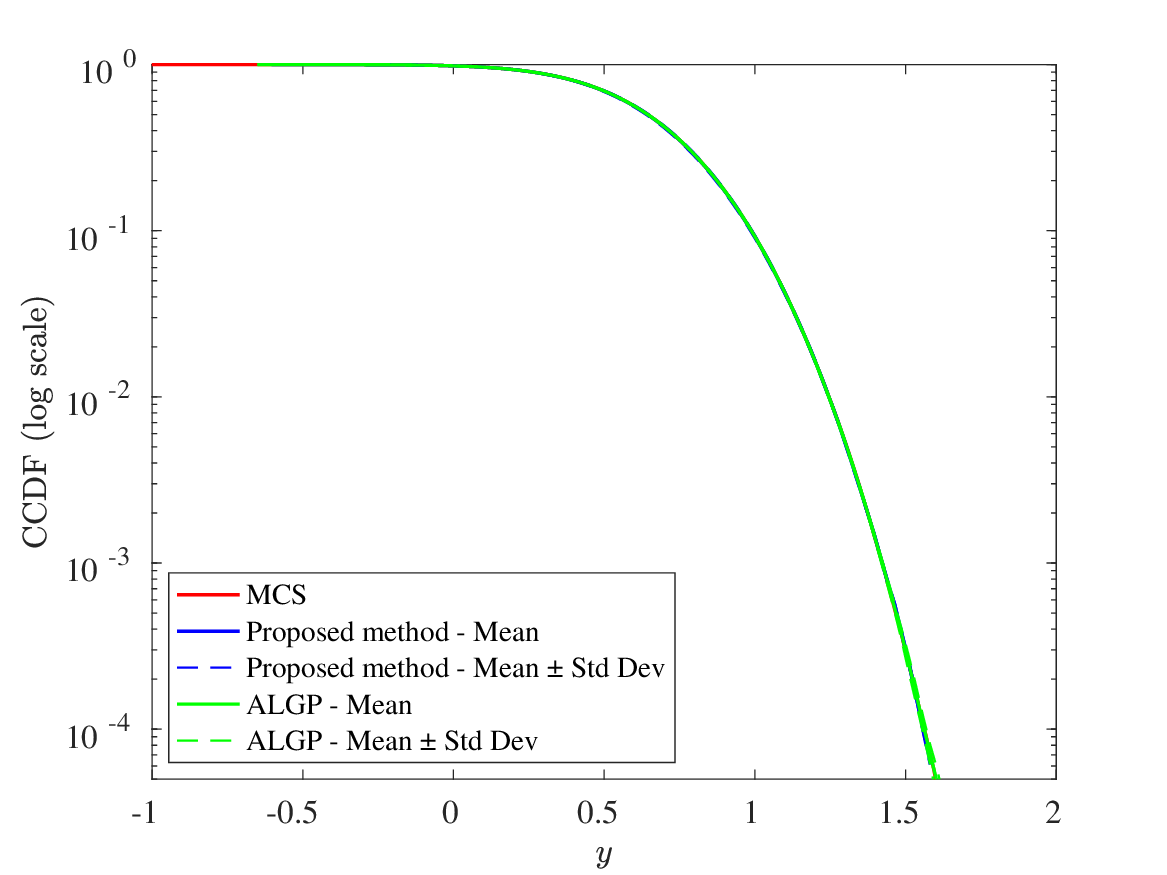}\label{fig:cdf_exam_3_b}
		\end{minipage}	
	}%
	
	\subfigure[Upper bound on the posterior CoV function of CDF]{
		\begin{minipage}{8.0cm}
			\centering
			\includegraphics[scale=0.35]{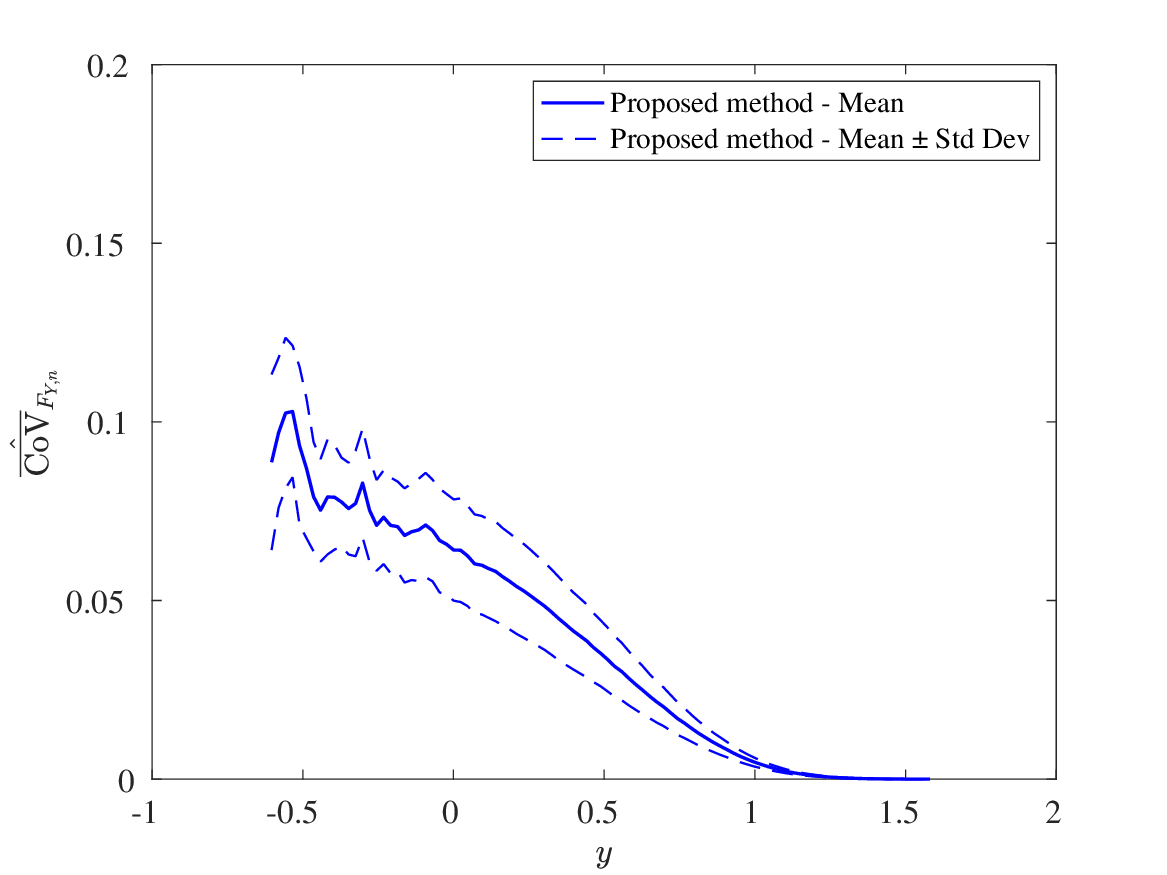}\label{fig:cdf_exam_3_c}
		\end{minipage}	
	}%
	\subfigure[Upper bound on the posterior CoV function of CCDF]{
		\begin{minipage}{8.0cm}
			\centering
			\includegraphics[scale=0.35]{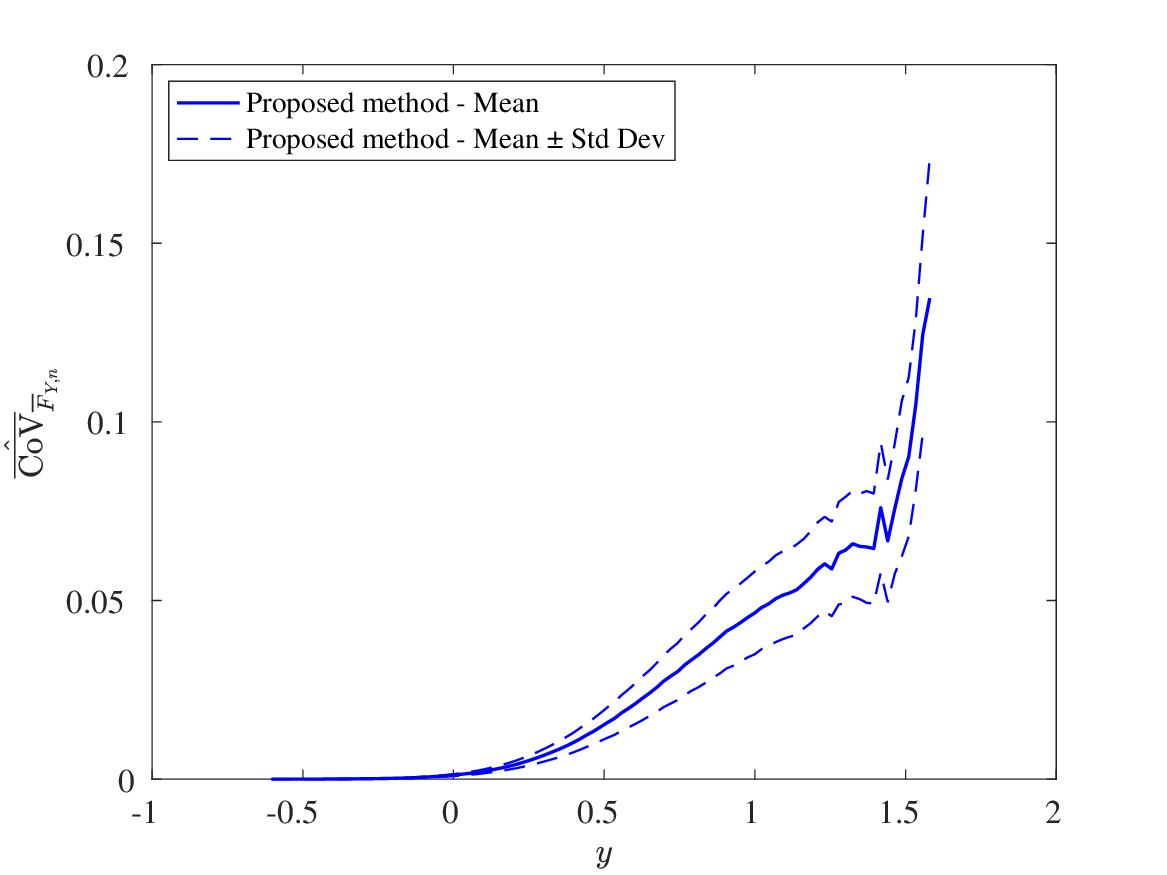}\label{fig:cdf_exam_3_d}
		\end{minipage}	
	}%
	
	\caption{Response CDF and CCDF  for Example 3.}
	\label{fig:CDF_exam_3}
\end{figure}

\begin{figure}[htb]
	\centering
	\includegraphics[scale=0.35]{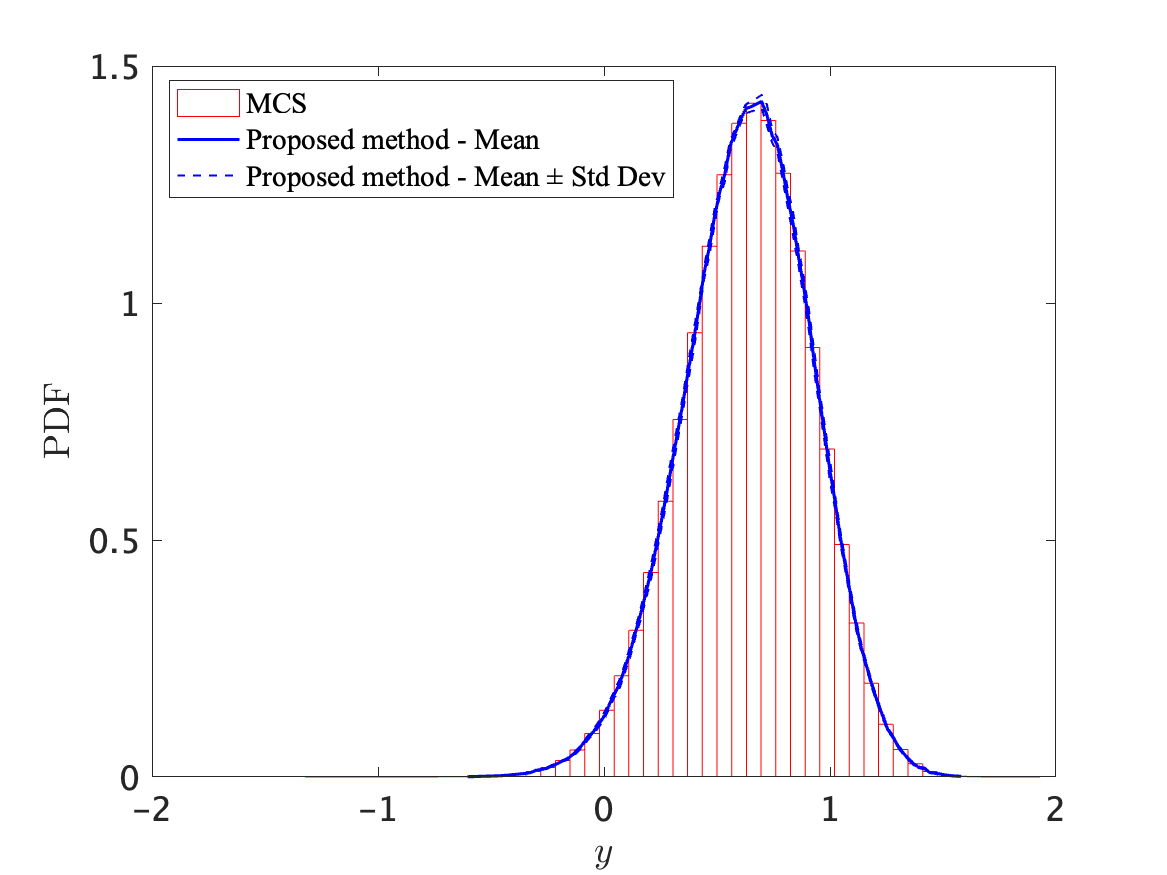}
	\caption{Response PDF  for Example 3.}	\label{fig:PDF_exam_3}
\end{figure}

To provide the reference results for the response CDF, CCDF and PDF,   MCS with $10^7$ samples is conducted.  For the AL-GP method, the range of interest and the stopping criterion are set to $[-0.65,1.65]$ and $\overline{\epsilon}=0.10$ respectively.  Figs. \ref{fig:cdf_exam_3_a} and \ref{fig:cdf_exam_3_b} demonstrate that both the AL-GP method and the proposed method are capable of generating CDF/CCDF mean curves that are in close alignment with the reference curves, as well as mean ± dev bounds that are notably narrow. The upper bounds of the posterior CoV functions for  the response CDF and CCDF are presented statistically in Figs. \ref{fig:cdf_exam_3_c} and \ref{fig:cdf_exam_3_d} respectively, as a result of  the proposed method.  Fig. \ref{fig:PDF_exam_3} shows the statistical results of the response PDF, produced by the proposed method as a by-product.  It can be observed that the PDF mean curve aligns closely with the histogram produced by MCS, and the mean ± dev band is relatively narrow.  Table \ref{tab:Ncall_exam_3} compares the required number of $g$-function calls. The proposed method, on average, necessitates a reduced number of $g$-function evaluations, while exhibiting a marginally greater CoV than the AL-GP method.

\begin{table}[htb]
	\centering
	\caption{Comparison of the number of $g$ function calls for Example 3.}\label{tab:Ncall_exam_3}
	\begin{tabular}{lll}
		\hline
		\multirow{2}{*}{Method}		&  \multicolumn{2}{c}{$N_{\mathrm{call}}$}  \\ \cline{2-3}
		
		&  Mean  &  CoV  \\
		\hline
		Proposed method &  10 + 14.75  = 24.75   & 8.68\%  \\
		AL-GP    &  12 + 21.60 =  33.60 &  7.87\%    \\
		\hline
	\end{tabular}
\end{table}

\subsection{Example 4: A space truss structure}

\begin{figure}[htb]
	\centering
	\includegraphics[scale=0.55]{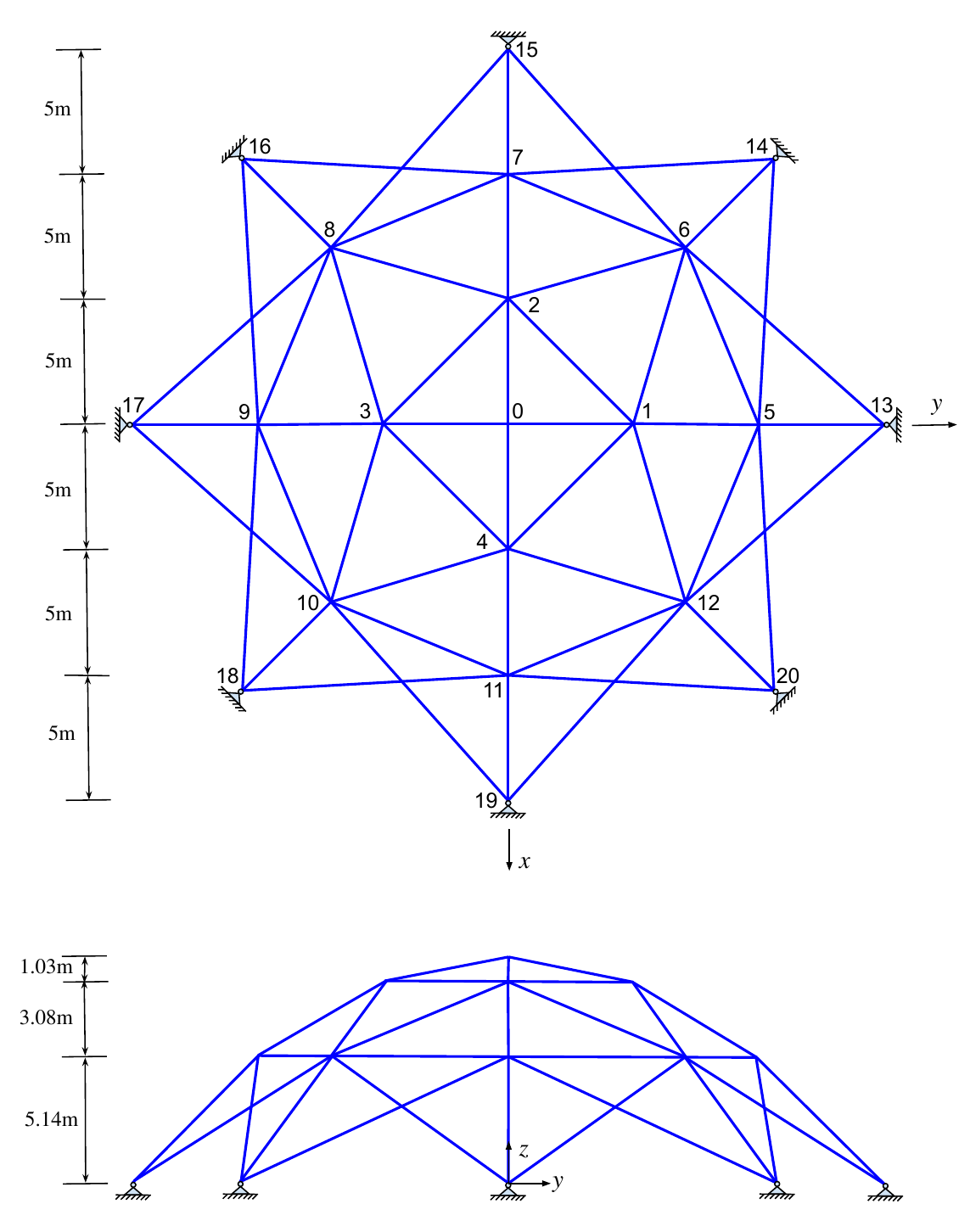}
	\caption{A 52-bar space truss structure.}\label{fig:truss}
\end{figure}

The fourth numerical example consists of a 52-bar space truss structure, as shown in Fig. \ref{fig:truss}. This structure is modeled as a three-dimensional finite element model using the open source software  OpenSees.  The model consists of 21 nodes and 52 truss elements. All elements have the same cross-sectional area $A$ and Young's modulus $E$. Concentrated vertical loads along the negative $z$-axis, denoted $P_0$ - $P_{12}$, are applied to nodes 0 - 12. Of interest is the vertical displacement of node 0:
\begin{equation}
     Y = g(A,E,P_{0}, P_{1},\cdots,P_{12} ),
\end{equation}
where $A$, $E$, $P_{0}$, $P_{1}$, $\cdots$, $P_{12} $ are treated as 15 independent random variables, as described in Table \ref{tab:rv_exam_4}.

\begin{table}[htb]
	\centering
    \caption{Input random variables for  Example 4.}\label{tab:rv_exam_4}
	\begin{tabular}{llll}
		 \hline
	Variable	 & Distribution  &  Mean &  CoV \\
	\hline
	      $A$              &           Normal             &        $2\times 10^3$ mm$^2$       &    0.10      \\
	    $E$              &           Normal             &        $2.06 \times 10^5$ MPa       &    0.10      \\  
	     $P_0$              &           Log-normal             &        $2 \times 10^2$ kN       &    0.20      \\  
	     $P_1, P_2, \cdots, P_{12}$              &           Log-normal             &        $1 \times 10^2$ kN       &    0.15      \\  
	\hline
	\end{tabular}
\end{table}

\begin{figure}[htb]
	\centering
	\subfigure[CDF]{
		\begin{minipage}{8.0cm}
			\centering
			\includegraphics[scale=0.35]{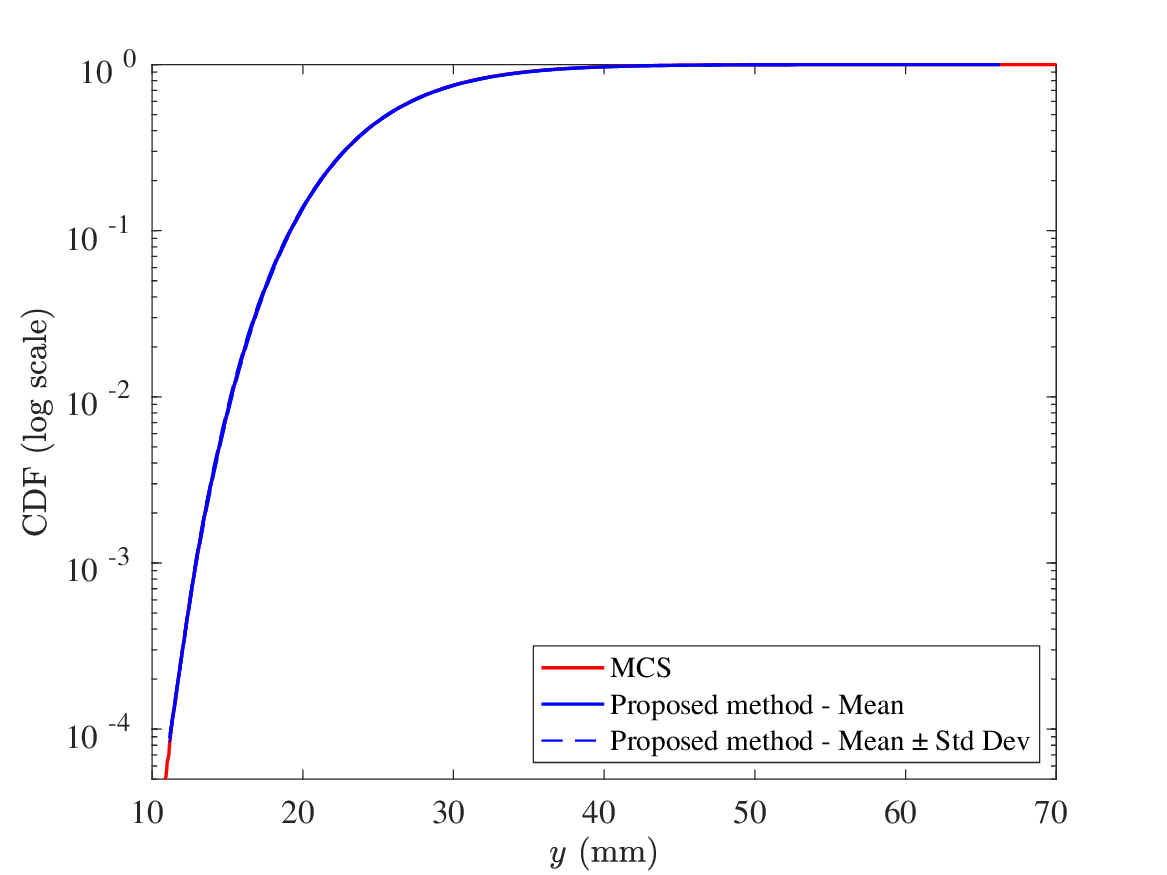}\label{fig:cdf_exam_4_a}
		\end{minipage}	
	}%
	\subfigure[CCDF]{
		\begin{minipage}{8.0cm}
			\centering
			\includegraphics[scale=0.35]{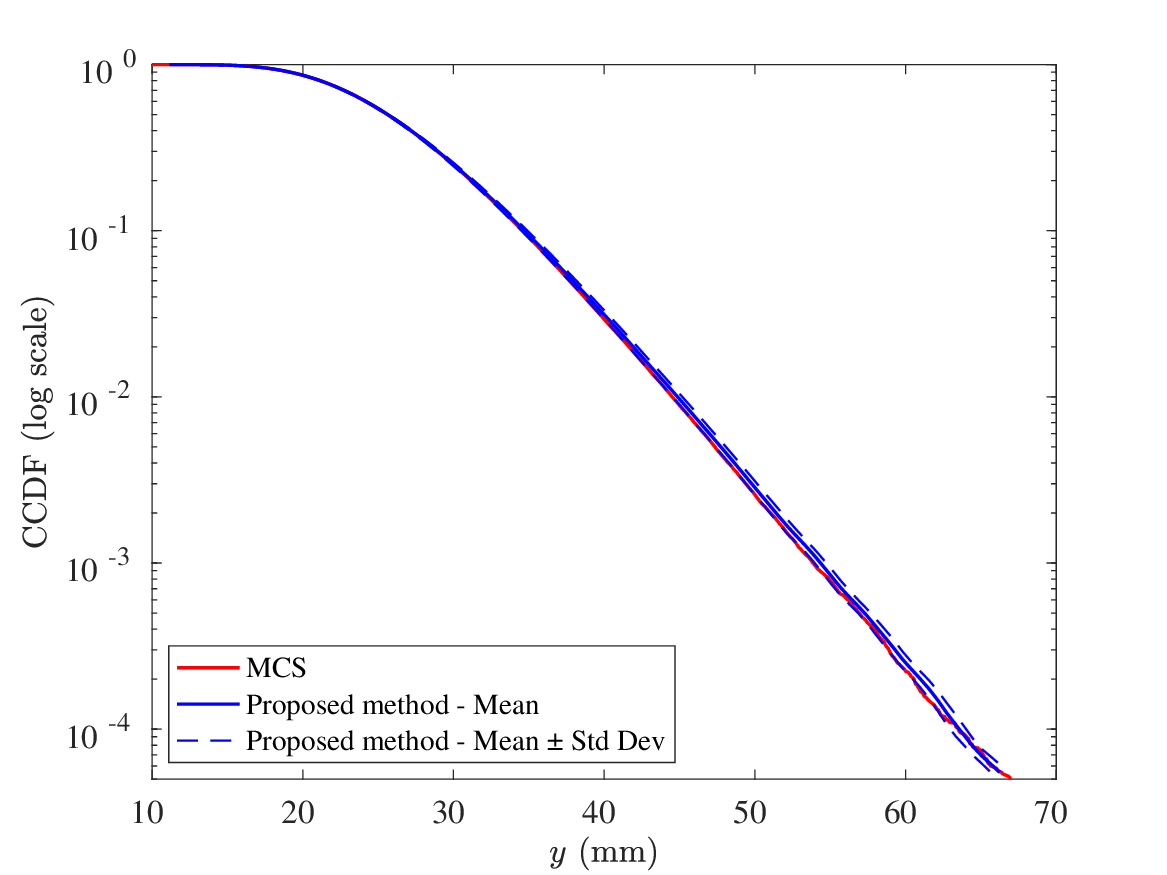}\label{fig:cdf_exam_4_b}
		\end{minipage}	
	}%
	
	\subfigure[Upper bound on the posterior CoV function of CDF]{
		\begin{minipage}{8.0cm}
			\centering
			\includegraphics[scale=0.35]{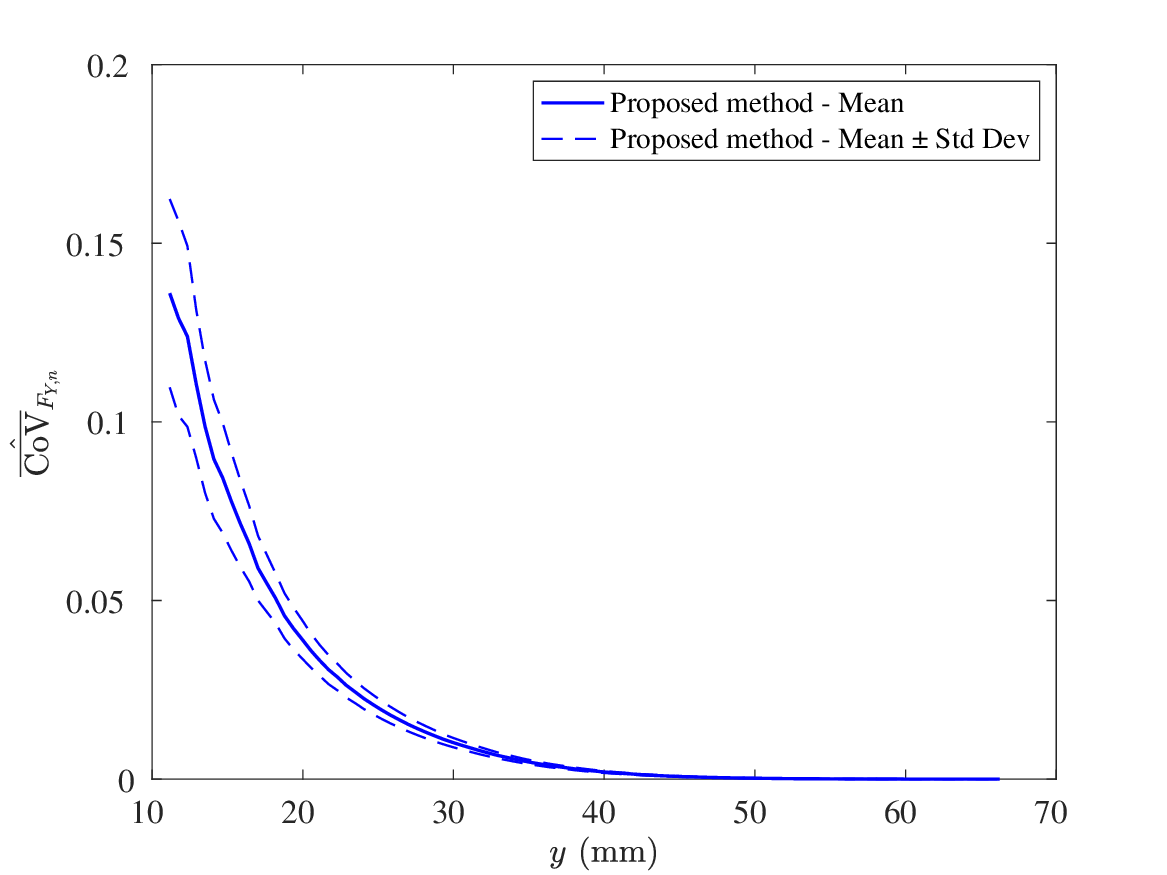}\label{fig:cdf_exam_4_c}
		\end{minipage}	
	}%
	\subfigure[Upper bound on the posterior CoV function of CCDF]{
		\begin{minipage}{8.0cm}
			\centering
			\includegraphics[scale=0.35]{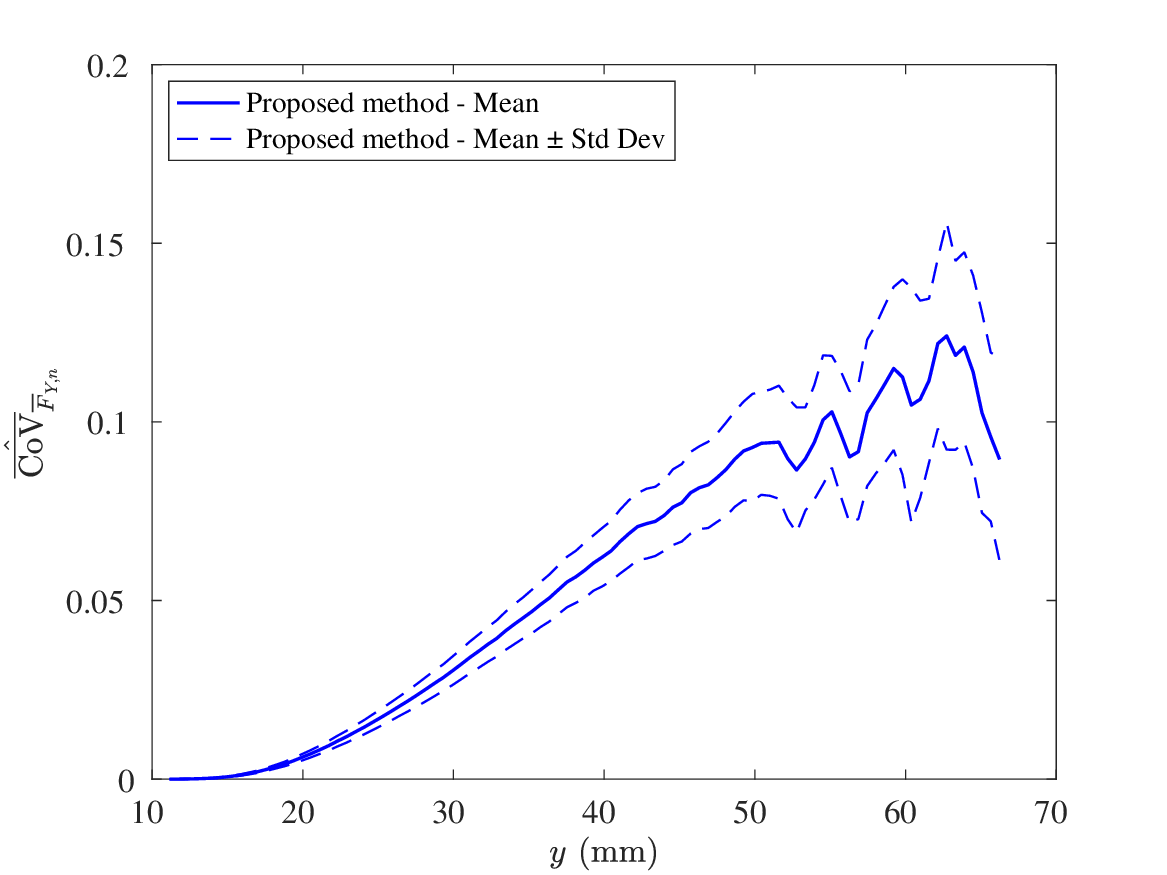}\label{fig:cdf_exam_4_d}
		\end{minipage}	
	}%
	
	\caption{Response CDF and CCDF  for Example 4.}
	\label{fig:CDF_exam_4}
\end{figure}

\begin{figure}[htb]
	\centering
	\includegraphics[scale=0.35]{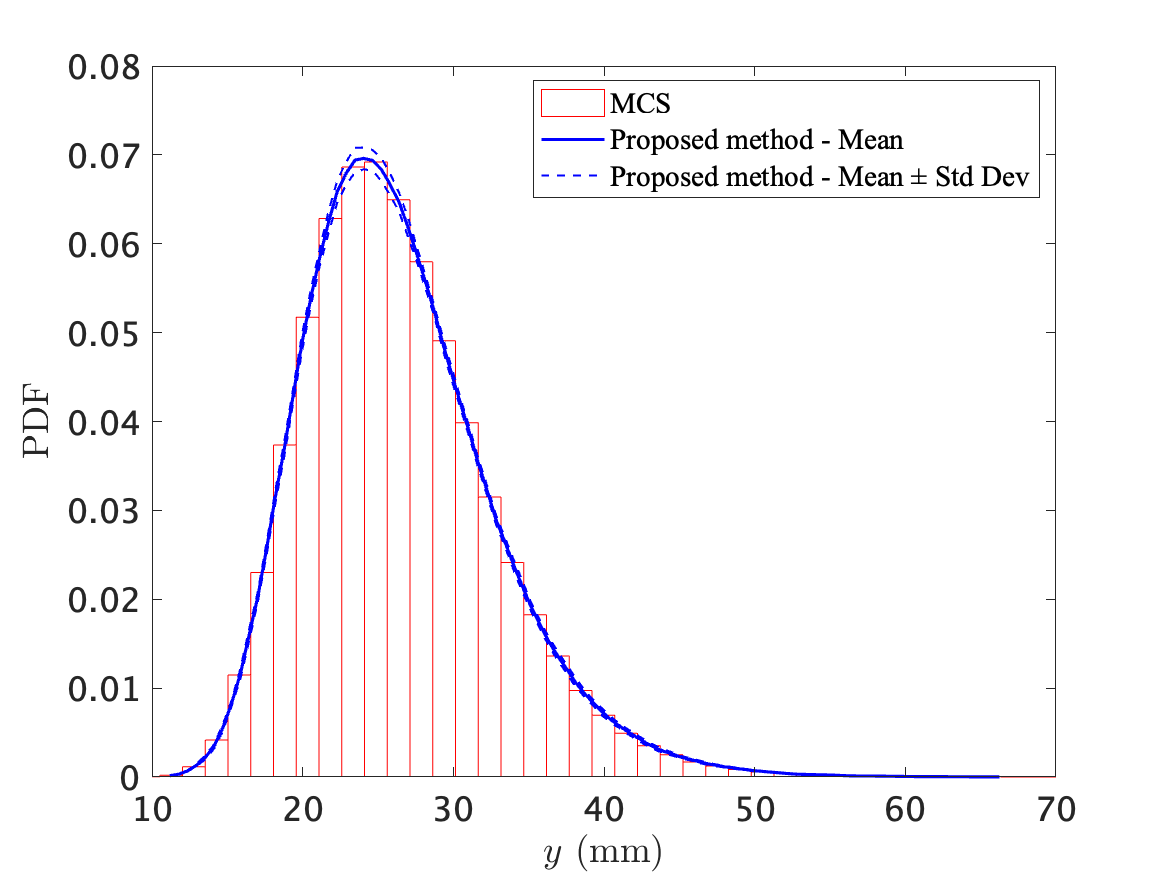}
	\caption{Response PDF  for Example 4.}\label{fig:PDF_exam_4}
\end{figure}

The reference solutions for the response CDF, CCDF and PDF are provided by MCS with $10^6$ samples.    Figs. \ref{fig:cdf_exam_4_a} and \ref{fig:cdf_exam_4_b}  shows that the proposed method can produce an almost unbiased CDF/CCDF mean and a narrow mean $\pm$ std dev band. In addition, the proposed method can also provide the local error measures, i.e. the upper bounds on the posterior CoV of the response CDF and CCDF, where the mean curves and mean $\pm$ std dev bands are plotted in Figs. \ref{fig:cdf_exam_4_c} and \ref{fig:cdf_exam_4_d}, respectively. The response PDF can be obtained as a by-product for the proposed method, the statistical results of which are shown in Fig. \ref{fig:PDF_exam_4}. Again, the results are very favorable. Note that the proposed method only requires an average of  10 + 29.50 = 39.50 $g$-function evaluations (with a CoV of 9.16\%).


%

\subsection{Example 5:  A seepage problem}

\begin{figure}[htb]
   \centering
   \includegraphics[scale=0.23]{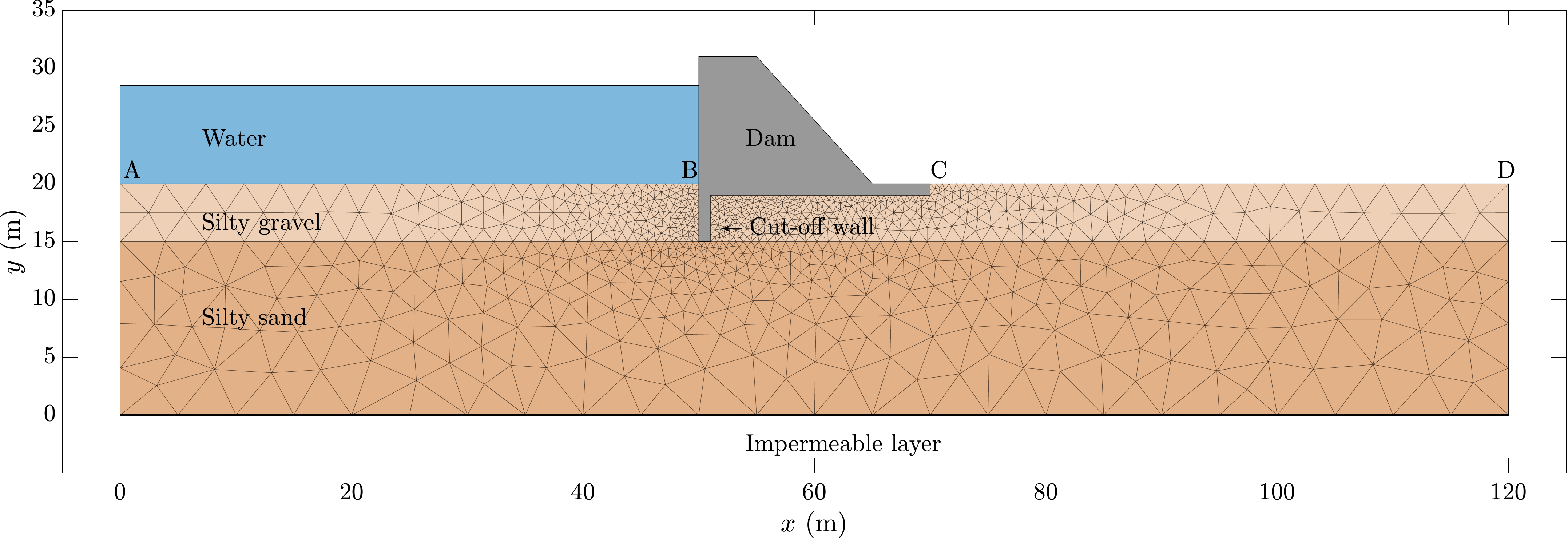}
   \caption{Schematic representation of the confined seepage  under an impermeable dam.}\label{fig:seepage_dam}
\end{figure}

\begin{figure}[htb]
	\centering
	\includegraphics[scale=0.23]{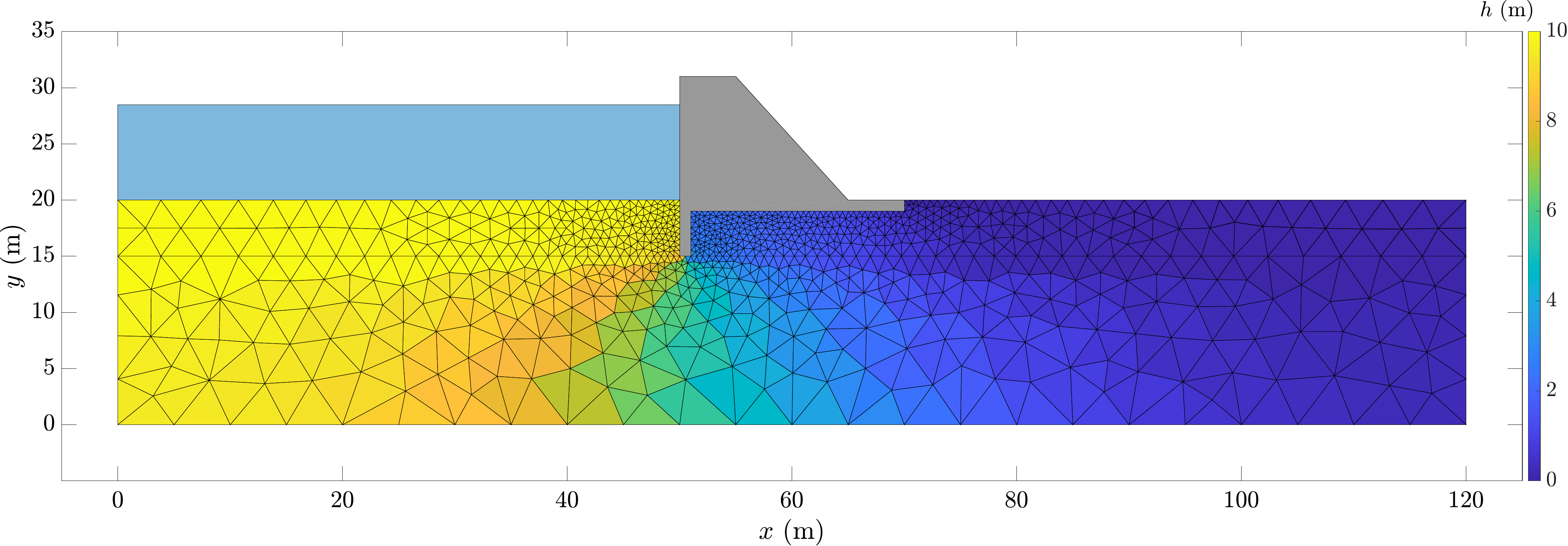}
	\caption{Hydraulic head solved by fixing the random variables at their mean values. }\label{fig:hydraulic}
\end{figure}

The last numerical example investigates the steady-state confined seepage under an impervious dam (adopted from \cite{valdebenito2018sensitivity}), as shown in Fig. \ref{fig:seepage_dam}.  The dam rests on an impermeable rock layer, above which are two permeable  soil layers: a 15 m  layer of silty sand and a 5 m  layer of silty gravel.  The  horizontal and vertical   permeabilities of the $i$-th layer are given as $k_{xx,i}$ and $k_{yy,i}$, respectively.  A water column of height $h_D$ m is retained at the upstream of the dam. The hydraulic head $h_W$ over the impermeable  layer is $h_W = h_D + 20$ m.  At the bottom of the dam, a cut-off wall is installed to prevent excessive seepage.   It is assumed that water flows only from segment AB to segment CD through the two permeable layers, with no flow occurring along any other boundary of the system.  We consider $h_D$, $k_{xx,1}$, $k_{yy,1}$, $k_{xx,2}$ and $k_{yy,2}$ as five independent random variables, as listed in Table \ref{tab:random_r_exam_5}. The hydraulic head $h_W$ is governed by the following partial differential equation: 
\begin{equation}\label{eq:pde}
   k_{x x, i} \frac{\partial^2 h_W}{\partial x^2}+k_{y y, i} \frac{\partial^2 h_W}{\partial y^2}=0, i=1,2,
\end{equation}
where $x$ and $y$ are the horizontal and vertical coordinates, respectively.   The boundary conditions for this equation include the hydraulic head across segments AB and CD, with zero flow across the remaining boundaries, as previously described.  Eq. \eqref{eq:pde} is solved using the finite element method with 3413 nodes and 1628 quadratic triangular elements, as shown in Fig. \ref{fig:seepage_dam}.  For illustrative purposes, the hydraulic head solved by fixing the random variables at their mean values is shown in Fig.  \ref{fig:hydraulic}.  Once the hydraulic head is solved, the quantity of interest, i.e. the seepage flow $q$ on the downstream side of the dam, can be calculated:
\begin{equation}
	q=-\int_{\mathrm{CD}} k_{y y, 2} \frac{\partial h_W}{\partial y} d x, 
\end{equation}
which is measured in units of volume over time over distance.  The flow $q$ is a function of the five basic random variables, so it is a random variable as well.

\begin{table}[htb]
  \centering
  \caption{Basic random variables for Example 5.}\label{tab:random_r_exam_5}
  \begin{threeparttable}
  	
  \begin{tabular}{llll}
       \hline
    Variable   &  Distribution  &  Parameter 1     &    Parameter 2   \\
    \hline
    $h_D ~(\rm m)$          &   Uniform    &      7                        &         10         \\
    $k_{xx,1}~(10^{-7} ~\rm m/s) $   &  Log-normal &     5              &   0.20             \\
    $k_{yy,1}~(10^{-7} ~\rm m/s) $   &  Log-normal &     2              &   0.20             \\
   $k_{xx,2}~(10^{-6} ~\rm m/s) $   &  Log-normal &     5              &   0.20             \\
   $k_{yy,2}~(10^{-6} ~\rm m/s) $   &  Log-normal &     2              &   0.20             \\
   \hline
  \end{tabular}
    \begin{tablenotes}
       \item Note: For a uniform distribution, parameter 1 and parameter 2 are the lower and upper bounds, respectively; otherwise, parameter 1 and parameter 2 are the mean and CoV, respectively. 
     \end{tablenotes}   	
\end{threeparttable}
\end{table}

\begin{figure}[htb]
	\centering
	\subfigure[CDF]{
		\begin{minipage}{8.0cm}
			\centering
			\includegraphics[scale=0.35]{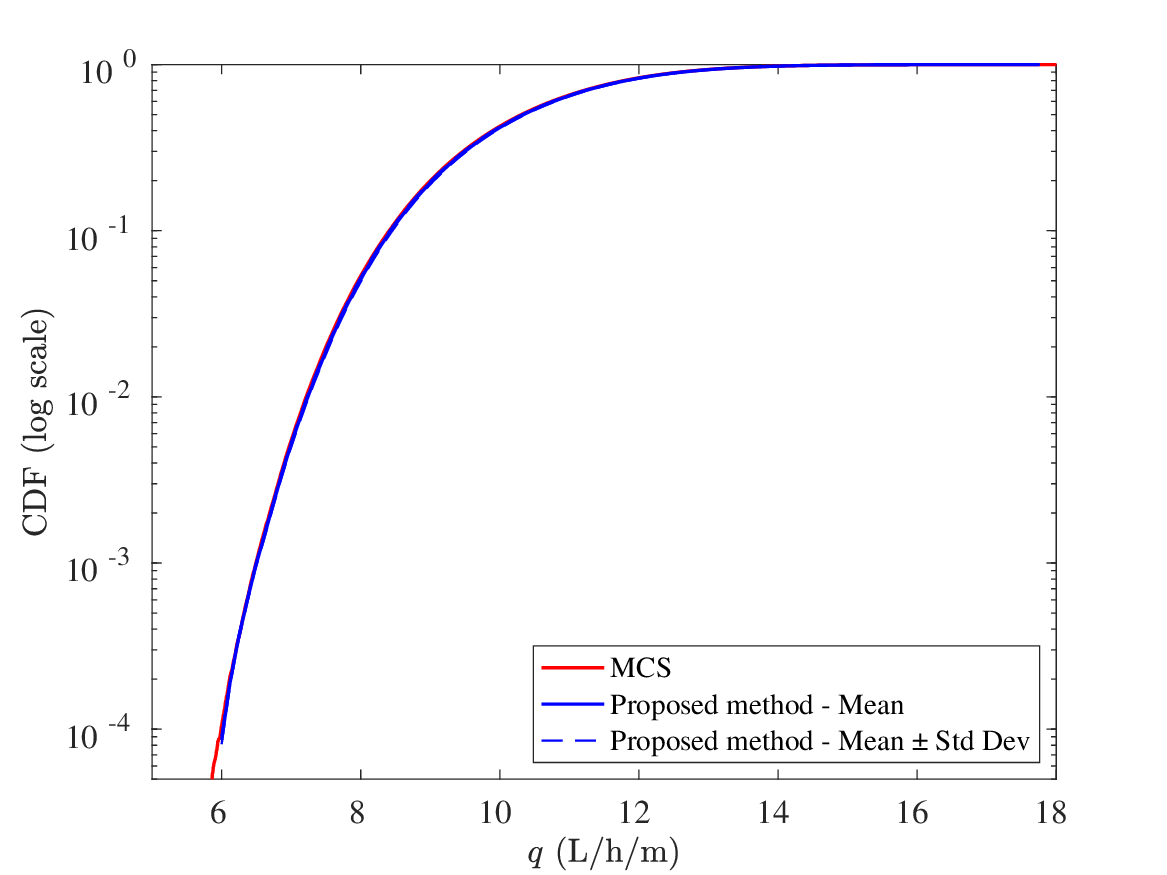}\label{fig:cdf_exam_5_a}
		\end{minipage}	
	}%
	\subfigure[CCDF]{
		\begin{minipage}{8.0cm}
			\centering
			\includegraphics[scale=0.35]{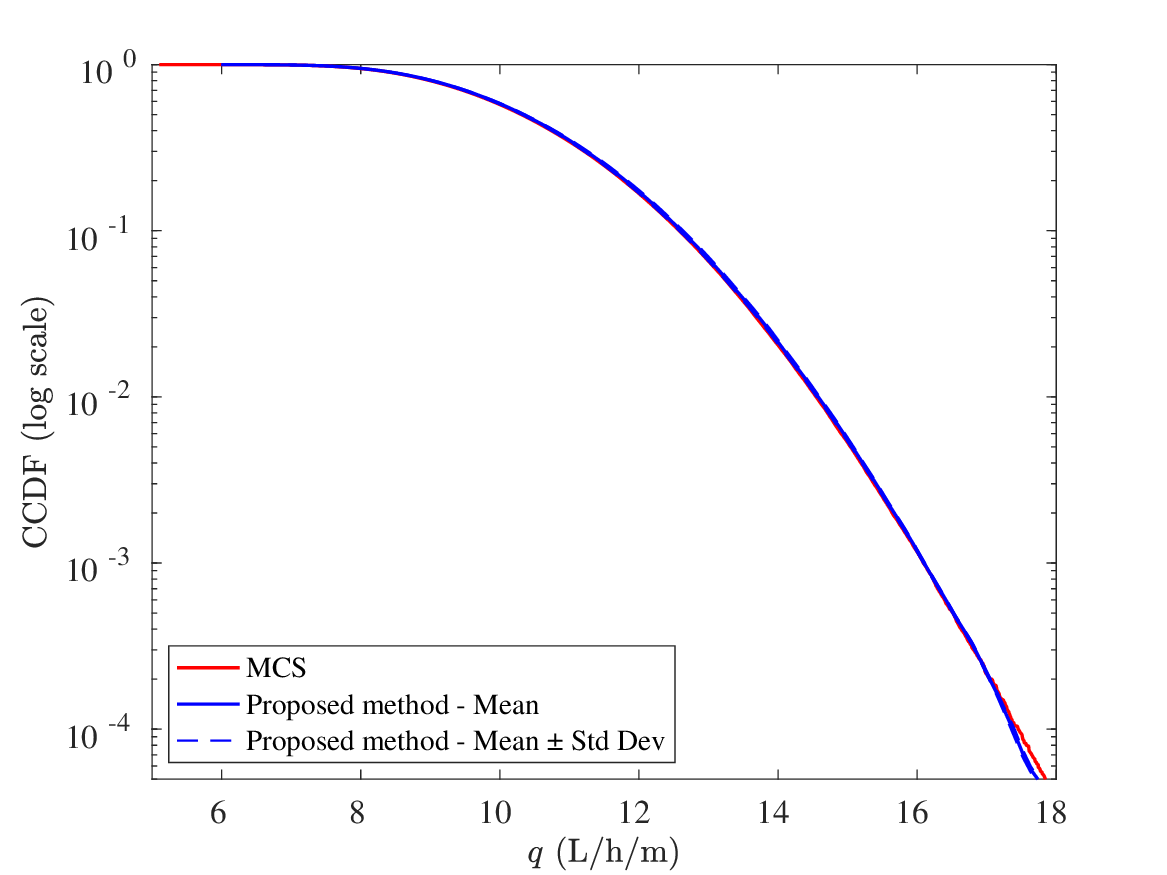}\label{fig:cdf_exam_5_b}
		\end{minipage}	
	}%
	
	\subfigure[Upper bound on the posterior CoV function of CDF]{
		\begin{minipage}{8.0cm}
			\centering
			\includegraphics[scale=0.35]{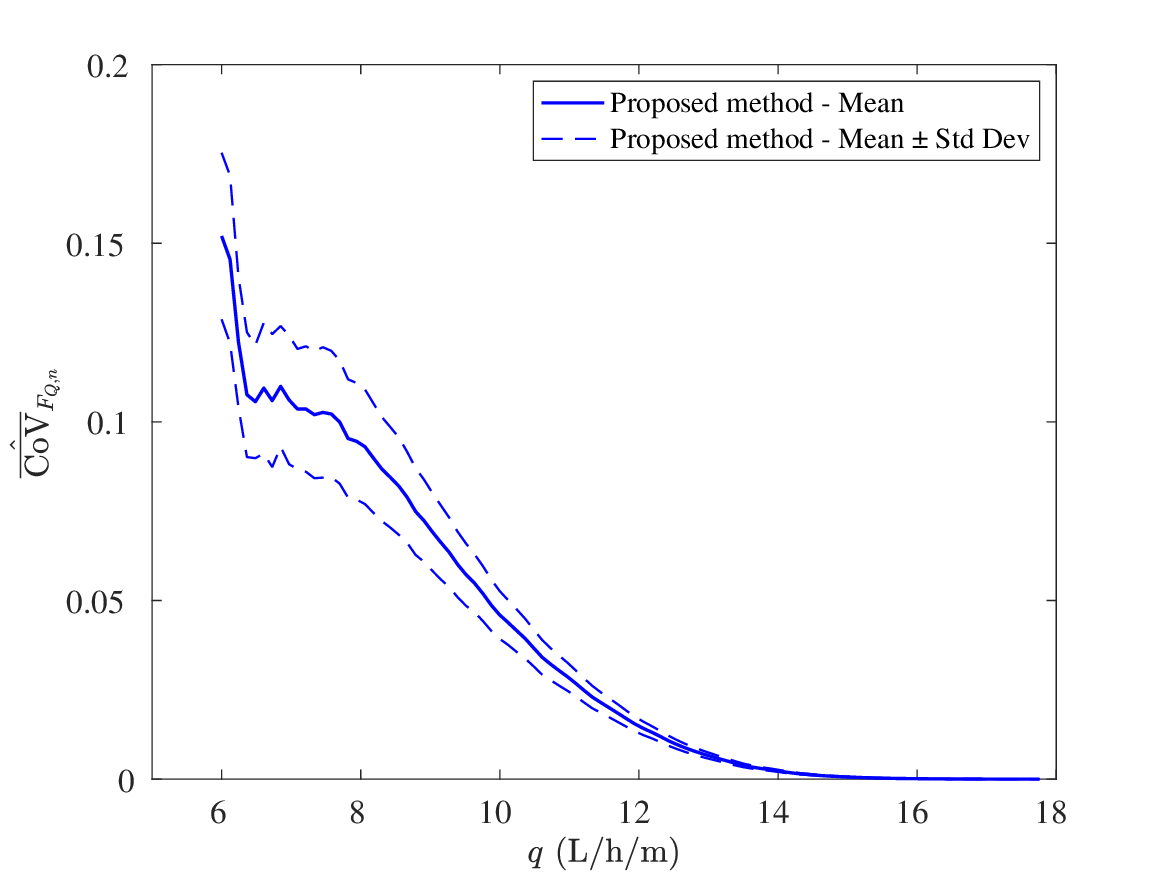}\label{fig:cdf_exam_5_c}
		\end{minipage}	
	}%
	\subfigure[Upper bound on the posterior CoV function of CCDF]{
		\begin{minipage}{8.0cm}
			\centering
			\includegraphics[scale=0.35]{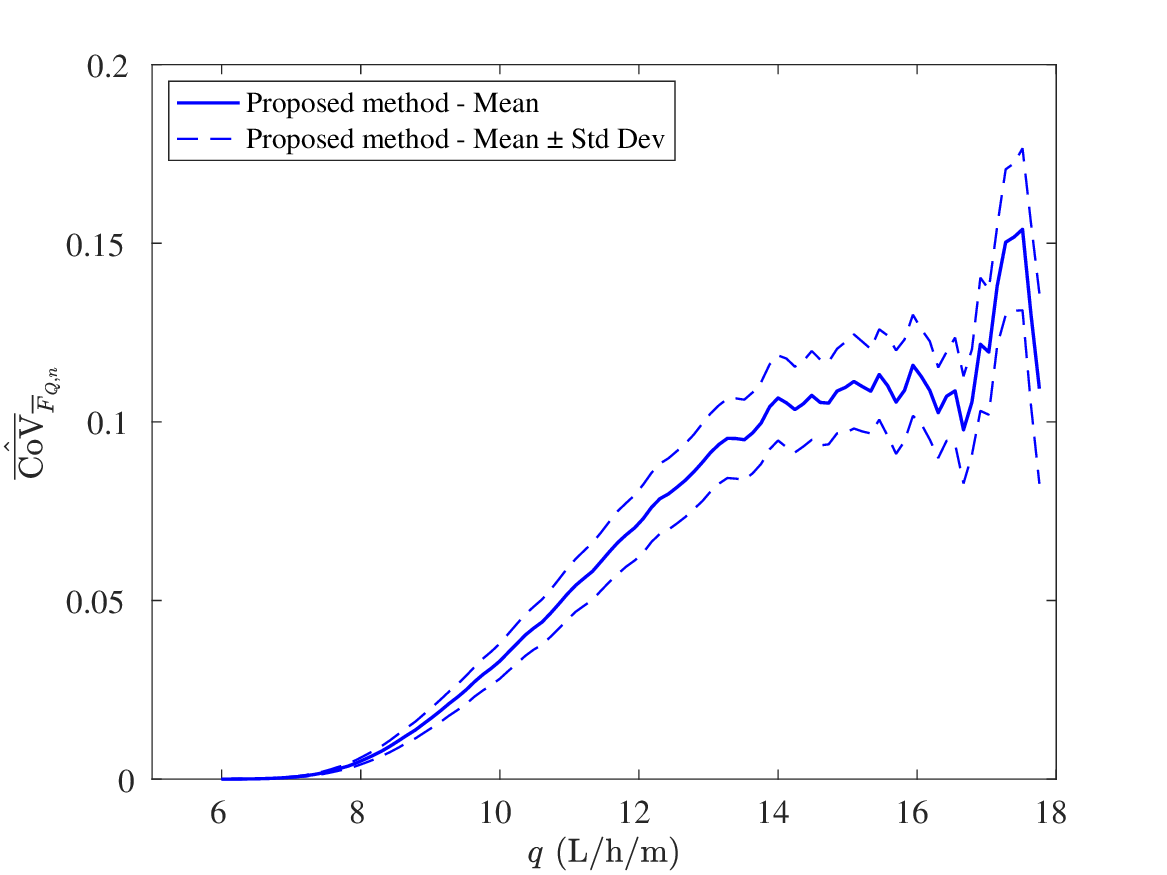}\label{fig:cdf_exam_5_d}
		\end{minipage}	
	}%
	
	\caption{Response CDF and CCDF  for Example 5.}
	\label{fig:CDF_exam_5}
\end{figure}

\begin{figure}[htb]
	\centering
	\includegraphics[scale=0.35]{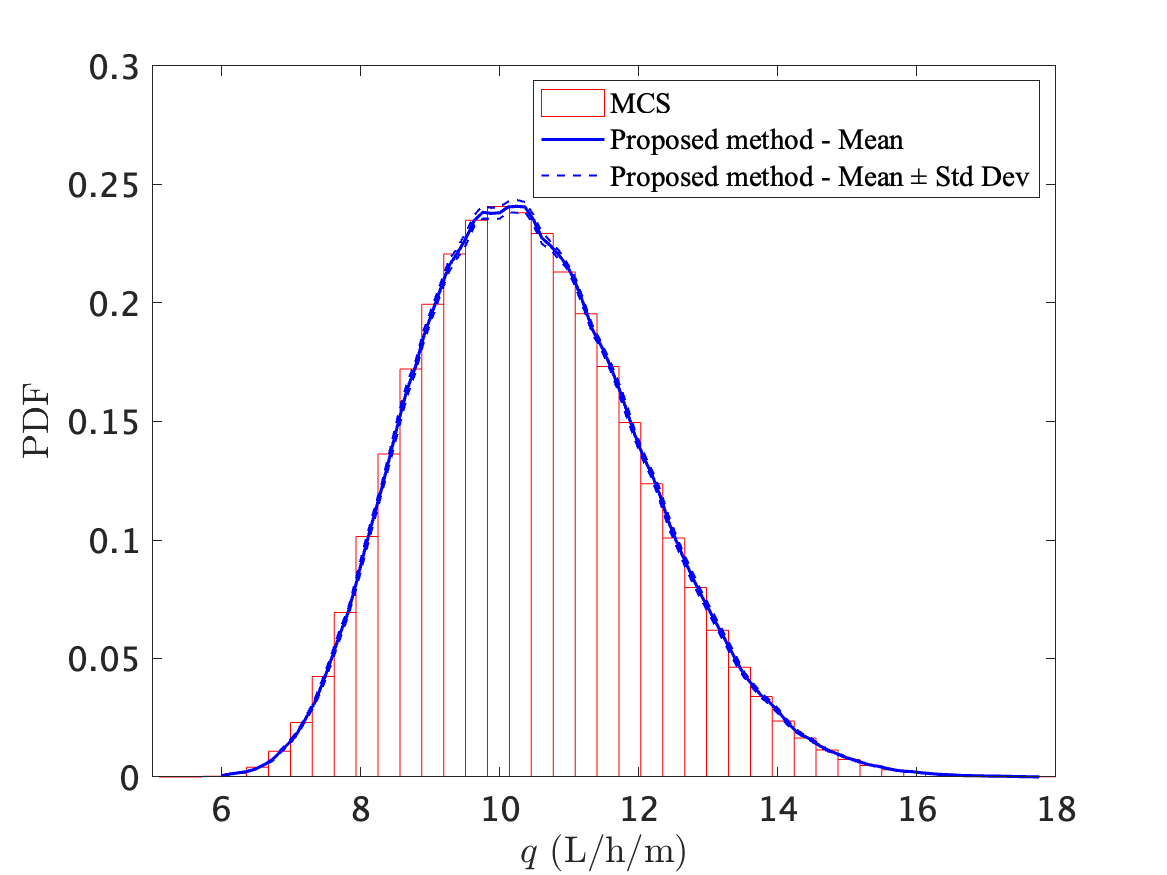}
	\caption{Response PDF  for Example 5.}\label{fig:PDF_exam_5}
\end{figure}

The reference results for the response CDF, CCDF, and PDF of $q$ are generated by MCS with  $10^6$ runs.  The CDF and CCDF related results from the proposed BAL method are shown in Fig.  \ref{fig:CDF_exam_5}. It can be seen from Figs.  \ref{fig:cdf_exam_5_a} and \ref{fig:cdf_exam_5_b} that  the proposed method produces CDF and CCDF mean curves that are close to the corresponding reference solutions, with very narrow mean ± std dev bands. Our method can also provide the upper bounds on the posterior CoV of the CDF and CCDF, whose statistical results  are depicted in Figs.  \ref{fig:cdf_exam_5_c} and \ref{fig:cdf_exam_5_d}.  In addition, the response PDF can be obtained as a by-product, as shown in Fig. \ref{fig:PDF_exam_5}, with the mean curve close to the reference result and a narrow mean ± std dev band. Remarkably, the proposed method only needs on average 10 + 24.65 = 34.65 model evaluations (with a CoV of 8.44\%).


%

\clearpage
\section{Concluding remarks}\label{sec:conclusions}

This paper presents a Bayesian active learning perspective using Gaussian process (GP) regression  on estimating the response probability distributions of expensive computer simulators in the presence of randomness. First, the estimation of response probability distributions is conceptually interpreted as a Bayesian inference problem, in contrast to   frequentist inference. This conceptual Bayesian idea  has several important advantages, for example,  quantifying numerical error  as a source of epistemic uncertainty, incorporating prior knowledge and  enabling the reduction of numerical error in an active learning way. By virtue of the well-established GP regression, a practical Bayesian approach is then developed for estimating the response probability distributions. In this context, a GP prior is assigned over the computer simulator, conditioning this GP prior on several computer simulator evaluations gives rise to a GP posterior for the computer simulator.  We  derive the posterior statistics of the response cumulative distribution function  (CDF), complementary CDF (CCDF) and probability density function (PDF) in semi-analytical form, and provide the numerical solution scheme. At last, a Bayesian active learning method is proposed for response probability distribution estimation, where two key components for active learning are devised by making use of the posterior statistics. Five numerical examples are studied to demonstrate the performance of the proposed Bayesian active learning method. The results show that our method can produce the response CDF and CCDF with quantified uncertainty using only a small number of calls to the computer simulator. Additionally, the method provides the response PDF as a by-product without numerically differentiating the response CDF.

The results of this study can be used as a starting point for a number of future studies. In particular, two possible research directions are suggested here.  One is to develop a Bayesian active learning scheme that can operate directly on the response PDF, which remains challenging.  The other is to develop a strategy that can identify multiple informative points from the learning function, thus allowing parallel computation.

\section*{Acknowledgments}

Chao Dang is grateful for the financial support of the German Research Foundation (DFG) (Grant number 530326817). Jun Xu would like to appreciate the financial support of National Natural Science Foundation of China (No. 52278178).

\section*{Data availability}
No data was used for the research described in the article.



  \bibliographystyle{elsarticle-num} 
  \bibliography{Reference.bib}





\end{document}